\documentclass[twocolumn]{aastex631}
\usepackage[T1]{fontenc}
\usepackage{textcomp}
\usepackage{xspace}
\usepackage{xcolor}

\newcommand\kerrc{{\tt KerrC \xspace}}

\newcommand{\comment}[1]{}

\newcommand{\wchi}{7136}
\newcommand{\dof}{1582}
\newcommand{\wa}{0.861}
\newcommand{\wMdot}{0.158}
\newcommand{\wtauC}{0.256}
\newcommand{\wtempC}{100} 
\newcommand{\wthetaC}{21.8} 
\newcommand{\wrCa}{72.9} 
\newcommand{\wkte}{1} 
\newcommand{\wlogNe}{20.8}

\newcommand{\fchi}{8001}
\newcommand{\fa}{0.921}
\newcommand{\fMdot}{0.167}
\newcommand{\ftauC}{0.64}
\newcommand{\ftempC}{107} 
\newcommand{\fthetaC}{45} 
\newcommand{\frCa}{41.4} 
\newcommand{\frCb}{89.3} 
\newcommand{\fkte}{50} 
\newcommand{\flogNe}{16.3}

\newcommand{\config}{68,040}

\received{tbd, 2022}
\revised{tbd, 2022}
\usepackage{amsmath}
\shorttitle{A new X-ray fitting model: kerrC}
\shortauthors{Krawczynski and Beheshtipour}

\begin{document}

\def\reference@jnl#1{{\it #1,}}

\def\aj{\reference@jnl{AJ}}                   
\def\actaa{\reference@jnl{Acta Astron.}}      
\def\araa{\reference@jnl{ARA\&A}}             
\def\apj{\reference@jnl{ApJ}}                 
\def\apjl{\reference@jnl{ApJ}}                
\def\apjs{\reference@jnl{ApJS}}               
\def\ao{\reference@jnl{Appl.~Opt.}}           
\def\apss{\reference@jnl{Ap\&SS}}             
\def\aap{\reference@jnl{A\&A}}                
\def\aapr{\reference@jnl{A\&A~Rev.}}          
\def\aaps{\reference@jnl{A\&AS}}              
\def\azh{\reference@jnl{AZh}}                 
\def\baas{\reference@jnl{BAAS}}               
\def\bac{\reference@jnl{Bull. astr. Inst. Czechosl.}}
\def\caa{\reference@jnl{Chinese Astron. Astrophys.}}
\def\cjaa{\reference@jnl{Chinese J. Astron. Astrophys.}}
\def\icarus{\reference@jnl{Icarus}}           
\def\jcap{\reference@jnl{J. Cosmology Astropart. Phys.}}
\def\jrasc{\reference@jnl{JRASC}}             
\def\memras{\reference@jnl{MmRAS}}            
\def\mnras{\reference@jnl{MNRAS}}             
\def\na{\reference@jnl{New A}}                
\def\nar{\reference@jnl{New A Rev.}}          
\def\pra{\reference@jnl{Phys.~Rev.~A}}        
\def\prb{\reference@jnl{Phys.~Rev.~B}}        
\def\prc{\reference@jnl{Phys.~Rev.~C}}        
\def\prd{\reference@jnl{Phys.~Rev.~D}}        
\def\pre{\reference@jnl{Phys.~Rev.~E}}        
\def\prl{\reference@jnl{Phys.~Rev.~Lett.}}    
\def\pasa{\reference@jnl{PASA}}               
\def\pasp{\reference@jnl{PASP}}               
\def\pasj{\reference@jnl{PASJ}}               
\def\rmxaa{\reference@jnl{Rev. Mexicana Astron. Astrofis.}}%
\def\qjras{\reference@jnl{QJRAS}}             
\def\skytel{\reference@jnl{S\&T}}             
\def\solphys{\reference@jnl{Sol.~Phys.}}      
\def\sovast{\reference@jnl{Soviet~Ast.}}      
\def\ssr{\reference@jnl{Space~Sci.~Rev.}}     
\def\zap{\reference@jnl{ZAp}}                 
\def\nat{\reference@jnl{Nature}}              
\def\iaucirc{\reference@jnl{IAU~Circ.}}       
\def\aplett{\reference@jnl{Astrophys.~Lett.}} 
\def\apspr{\reference@jnl{Astrophys.~Space~Phys.~Res.}}
\def\bain{\reference@jnl{Bull.~Astron.~Inst.~Netherlands}} 
\def\fcp{\reference@jnl{Fund.~Cosmic~Phys.}}  
\def\gca{\reference@jnl{Geochim.~Cosmochim.~Acta}}   
\def\grl{\reference@jnl{Geophys.~Res.~Lett.}} 
\def\jcp{\reference@jnl{J.~Chem.~Phys.}}      
\def\jgr{\reference@jnl{J.~Geophys.~Res.}}    
\def\jqsrt{\reference@jnl{J.~Quant.~Spec.~Radiat.~Transf.}}
\def\memsai{\reference@jnl{Mem.~Soc.~Astron.~Italiana}}
\def\nphysa{\reference@jnl{Nucl.~Phys.~A}}   
\def\physrep{\reference@jnl{Phys.~Rep.}}   
\def\physscr{\reference@jnl{Phys.~Scr}}   
\def\planss{\reference@jnl{Planet.~Space~Sci.}}   
\def\procspie{\reference@jnl{Proc.~SPIE}}   

\let\astap=\aap
\let\apjlett=\apjl
\let\apjsupp=\apjs
\let\applopt=\ao
\title{New Constraints on the Spin  
of the Black Hole Cygnus X-1 and the Physical 
Properties of its Accretion Disk Corona}
\author{H.\,Krawczynski} \affil{Washington University in St. Louis, 
Physics Department, 
McDonnell Center for the Space Sciences, and 
the Center for Quantum Sensors,
1 Brookings Dr., CB 1105, St. Louis, MO 63130} 
\author{B.\,Beheshtipour} \affil{Max Planck Institute for 
Gravitational Physics, Albert Einstein Institute, Callinstrasse 38, 30167 Hannover, Germany}
\affil{Leibniz Universit\"at Hannover, D-30167 Hannover, Germany}
\correspondingauthor{Henric Krawczynski, krawcz@wustl.edu}

\begin{abstract}
We present a new analysis of {\it NuSTAR} and {\it Suzaku} observations of the black hole Cygnus X-1 in the intermediate state. The analysis uses {\tt kerrC}, a new model for analyzing spectral and spectropolarimetric X-ray observations of black holes. {\tt kerrC} builds on a large library of simulated black holes in X-ray binaries.  The model accounts for the X-ray emission from a geometrically thin, optically thick accretion disk, the propagation of the X-rays through the curved black hole spacetime, the reflection off the accretion disk, and the Comptonization of photons in coronae of different 3-D shapes and physical properties before and after the reflection. We present the results from using {\tt kerrC} for the analysis of archival {\it NuSTAR} and {\it Suzaku} observations taken on May 27-28, 2015. The best wedge-shaped corona gives a better fit than the cone-shaped corona. Although we included cone-shaped coronae in the funnel regions above and below the black hole to resemble to some degree the common assumption of a compact lamppost corona hovering above and/or below the black hole, the fit chooses a very large version of this corona
that makes it possible to Comptonize a sufficiently
large fraction of the accretion disk photons to
explain the observed power law emission.
The analysis indicates a black hole spin parameter $a$ ($-1 \le a \le  1$) between \wa\,and \fa. The {\tt kerrC} model provides new insights about the radial distribution of the energy flux of returning and coronal emission irradiating the accretion disk. {\tt kerrC} furthermore predicts small polarization fractions around 1\% in the 2-8 keV energy range of the recently launched {\it Imaging X-ray Polarimetry Explorer}.
\end{abstract}


\section{Introduction}
The 2020-2030 decade promises to lead to several 
breakthrough discoveries in the field of 
astrophysical studies of black holes and black hole accretion.
The pointed 
{\it NuSTAR} \citep{2013ApJ...770..103H}, 
{\it NGO} \citep{2004ApJ...611.1005G}, 
{\it NICER} \citep{2016SPIE.9905E..1HG}, 
{\it Chandra} \citep{2019SPIE11116E..02H}, and
{\it XMM-Newton} \citep{2017xru..conf..114K} 
X-ray observatories will continue to operate, 
and will be joined by the X-ray and $\gamma$-ray missions 
{\it IXPE} \citep{2021arXiv211201269W}, 
{\it XL-Calibur} \citep{2021APh...12602529A}, 
and {\it COSI} \citep{2020ApJ...897...45S} 
with polarimetric capabilities, 
and by the {\it XRISM} mission \citep{2020SPIE11444E..22T} 
with unprecedented 
high-throughput high-spectral-resolution capabilities.
At the same time, numerical General Relativistic Magnetohydrodynamic (GRMHD) and two-temperature 
General Relativistic Radiation 
Magnetohydrodynamic (2tGRRMHD) simulations developed 
by several groups allow us 
to model black hole accretion from first principles with
continually improving fidelity, see 
\citep{2017MNRAS.466..705S,2019MNRAS.486.2873C,2019ApJS..243...26P,2021ApJ...922..270K,2021MNRAS.507..983L} and references therein

We report here on a new tool, called {\tt kerrC} that we developed to
bridge the gap between numerical simulations and observations.
{\tt kerrC} models the X-ray flux and polarization 
energy spectra (Stokes $I$, $Q$, and $U$) for the thermal, 
intermediate, and hard states. In these states, 
black holes are thought to accrete through a geometrically thin, 
optically thick accretion disk surrounded by hot coronal plasma \citep[e.g.,][]{2006ARA&A..44...49R,2014SSRv..183..295M}.
The accretion disk emits a multi-temperature thermal 
component peaking at keV-energies which follows roughly the analytical 
models of \citet{1973A&A....24..337S,1973blho.conf..343N}.    
The Comptonization of the emission in a hot corona is believed
to explain the non-thermal continuum emission extending from 
a few keV to a few ten keV or even a few hundred keV.
Returning and coronal emission irradiating the disk 
gives rise to a reflection component, including the kinematically
and gravitationally broadened fluorescent Fe K-$\alpha$ 
emission around 6.4\,keV  \citep{1989MNRAS.238..729F}.
Some X-ray spectroscopic observations show furthermore
evidence for winds \citep{Miller_2016}, and some 
hard X-ray polarization and  $\gamma$-ray observations 
suggest the presence of a collimated plasma outflow (jet)
\citep[e.g.,][]{2017MNRAS.471.3657Z}. 

We focus in the following on the disk, the coronal, and 
the reflected emission.  As these emission components 
originate close to the black hole, they can 
inform us about the extremely curved spacetime close to the 
event horizon of the black hole, 
and allow us to constrain the inclination 
(the angle between 
the spin axis and the observer) and the spin
of the black hole
\citep{2014SSRv..183..295M,2006AN....327..997M}.
As the fidelity of our models continues to improve, 
test of  general relativity based on X-ray observations
may become feasible \citep{2016CQGra..33l4001J,2018GReGr..50..100K,2021SSRv..217...65B}.
As the accretion flow dissipates most of its energy close 
to the black hole, the three emission components are furthermore 
excellent diagnostics of the accretion flow physics. 
They present us with the opportunity to test 
models of the vertical structure of the accretion disk, 
the angular momentum transport inside and outside of the disk, 
the transfer of energy between ions, electrons, 
magnetic field, and radiation,
and the properties of the accreted plasma
and magnetic field \citep[e.g.][]{2014ARA&A..52..529Y}.

Current state-of-the-art analyses \citep[e.g.][]{2018ApJ...855....3T} 
fit the energy spectra with
a combination of a multi-temperature disk model such as {\tt KERRBB} \citep{2005ApJS..157..335L} 
and {\tt DISKBB} \citep{1984PASJ...36..741M,1986ApJ...308..635M,1998PASJ...50..667K} plus
a powerlaw or Comptonized powerlaw \citep[e.g.,][]{1996MNRAS.283..193Z,1999MNRAS.309..561Z} 
and a model of the reflected emission.
The energy spectra of the reflected emission are calculated with models such 
as {\tt reflionx}, {\tt reflionx\_hd} \citep{2005MNRAS.358..211R,1999MNRAS.306..461R}, 
or {\tt XILLVER} \citep{2011ApJ...731..131G,2013ApJ...768..146G,2014ApJ...782...76G}
that are based on radiative transfer calculations in a plane parallel atmosphere. 
The emission components are convolved with a kernel to account for the frequency shifts 
from the relativistic motion of the emitting 
and reflecting plasma and the gravitational frequency shift incurred by the X-rays as they climb 
out of the curved Kerr spacetime \citep[e.g.][]{2010MNRAS.409.1534D}.

The {\tt RELXILL} model of \citet{2014MNRAS.444L.100D,2014ApJ...782...76G} 
assumes a point source of hard X-rays 
positioned at a height $h$ on the rotation axis of the black hole. 
The lamppost model predicts the dependence of the flux irradiating the accretion disk as a function of radial distance $r$, but cannot
predict the absolute coronal flux, 
the energy spectrum of the coronal emission, 
nor the polarization of the coronal emission. 
The model assumes that the coronal energy spectrum and the density 
of the reflecting plasma follow simple parameterizations.
The lamppost assumption makes it possible to account for the 
dependence of the energy spectrum of the reflected emission 
on the direction of the reflected photons.

{\tt kerrC} replaces the lamppost hypothesis with spatially extended coronae of different shapes and locations, and different physical properties
(electron temperature and density). Modeling the Comptonization of the 
accretion disk emission in the corona from first principles, 
{\tt kerrC} can predict the absolute flux of the Comptonized emission, and the
radial dependence of the flux and energy spectrum of the coronal 
emission irradiating the accretion disk. The modeling with {\tt kerrC} can 
be used to measure the black hole spin and inclination and to constrain 
the shape and physical properties of the corona.

{\tt kerrC} implements the following features:
\begin{itemize}
\item It includes a wide range of pre-calculated models:
we simulated black hole spin parameters $a$ between 
-1 (maximally rotating black hole with counter-rotating accretion disk) and +0.998 (near maximally spinning black hole with co-rotating accretion disk), for widely varying coronal geometries 
(wedge and cone-shaped coronae of different locations and sizes),
and physical parameters (electron densities and temperatures).
\item  {\tt kerrC} makes absolute rather than relative predictions of the thermal, Comptonized, and reflected emission. This allows for a comprehensive test of the underlying assumptions. 
For example, {\tt kerrC} can be used to evaluate if a corona can be very close 
to the black hole as indicated by some spectral and timing studies
\citep[e.g.][]{2001MNRAS.328L..27W,2009Natur.459..540F,2015MNRAS.446..759C,2014A&ARv..22...72U}, and,
at the same time be sufficiently large  to 
Comptonize enough accretion disk photons 
to account for the observed power law and reflected 
emission components \citep{2016AN....337..441D,2020A&A...644A.132U}.
Furthermore, fitting the continuum and the lines at the same time,
{\tt kerrC} can constrain system parameters such as
the black hole spin and inclination more tightly than models
that use additional fitting parameters.
\item The model tracks the thermal, coronal, and returning 
emission, and the emission reflected off the accretion disk. 
\item {\tt kerrc} accounts for the Comptonization of the emission following the reflection off the disk \citep[see the related discussion by][]{2017ApJ...836..119S}.
\item {\tt kerrC} uses pre-calculated geodescis which are convolved with the chosen 
reflection model ``on the fly'' while fitting the data. 
The architecture makes it possible to convolve the {\tt kerrC} configurations with
the {\tt XILLVER} models, and to introduce and fit additional parameters.
{\tt kerrC} allows for example to fit a parameter describing 
the dependence of the photospheric electron density on 
the distance from the black hole. 
\item {\tt kerrC} calculates the flux (Stokes $I$) and polarization
(Stokes $Q$ and $U$) energy spectra. The model can thus be used 
for the analysis and interpretation of the flux and 
polarization energy spectra acquired with the recently launched
{\it Imaging X-ray Polarimetry Explorer (IXPE)} mission and 
the upcoming X-ray polarization mission {\it XL-Calibur}. 
\end{itemize}

The approach described here is complimentary to parallel efforts 
that combine first-principle GRMHD or 2tGRRMHD simulations
with raytracing calculations
\citep[e.g.,][West, A., Liska, M., et al., in preparation]{2016ApJ...826...52K,2019ApJ...873...71K,2020ApJ...904..117K,2021ApJ...922..270K,2018MNRAS.474L..81L,2019MNRAS.487..550L,2019arXiv191210192L,2020MNRAS.494.3656L,2021MNRAS.507..983L}.
We envisage that the results from first-principle simulations can be used in future  
to refine the {\tt kerrC} model. Conversely, the results from fitting 
observational data with {\tt kerrC} can be used to identify
shortcomings of the first-principle simulations. 

The interested reader is referred to 
\citep{
1985A&A...143..374S,1994ApJ...432L..95H,1994scs..book.....N,1994JQSRT..51..813P,
1996ApJ...470..249P,1996MNRAS.283..193Z,1997MNRAS.292L..21P,1999MNRAS.309..561Z,
2010ApJ...712..908S,2013MNRAS.430.1694D,2015MNRAS.448..703W,
2017MNRAS.472.1932G, 2017ApJ...850...14B,
2019ApJ...875..148Z}
for earlier analytical and numerical studies of the properties of the coronal emission. 

The rest of the paper is organized as follows. We describe the numerical simulations in 
Sect.\,\ref{s:model}, and the {\tt kerrC} implementation in Sect\,\ref{s:kerrC}.
We present the results from fitting {\it Suzaku} and {\it NuSTAR} observations of Cyg\,X-1 in Sect.\,\ref{s:res}.
We present predictions for {\it IXPE} in Sect.\,\ref{s:ixpe}.
We summarize and discuss the results in Sect.\,\ref{s:disc}. 

In the following we use $c=1$, and define the gravitational radius of the black 
hole of mass $M$ to be $r_{\rm g}\,=\,G M$.  
\section{Numerical Simulations}
\label{s:model}
\subsection{Overall Architecture and Simulated Configurations}
The {\tt kerrC} fitting model is based on a library of  
\config~raytraced black hole, accretion flow, and corona configurations. For each configuration, 20 million events are generated. 
After describing the raytracing simulations in this section, 
we detail how the simulated data are used in the fitting code in Section \ref{s:kerrC}. 

\begin{deluxetable*}{|p{6cm}|c|c|p{8cm}|}
\tablenum{1}
\tablecaption{Simulated {\tt kerrC} black hole, accretion disk, and corona configurations.\label{t:conf}}
\tablewidth{0pt}
\tablehead{
\colhead{Parameter} & \colhead{Symbol} & \colhead{Unit} & \colhead{Simulation Grid}}
\startdata
\multicolumn{4}{|c|}{Black Hole Parameters}\\ \hline
Black Hole Spin Parameter & $a$ & none & 
-1,0, 0.5, 0.75, 0.9, 0.95, 0.98, 0.99, 0.998\\ \hline
\multicolumn{4}{|c|}{Accretion Flow Parameters}\\ \hline
Temperature Scaling Factor &  $\sigma$  & none & 0.05, 0.1, 0.25, 0.5, 0.75, 1, 1.25, 1.5, 1.75, 2 \\ \hline
\multicolumn{4}{|c|}{Corona Parameters}\\ \hline
Corona Geometry &  $g$  &  none & 0 (wedge-shaped corona), 1 (cone-shaped corona) \\
Wedge Radius (Geom. \#1) & $r_C$ & $r_{\rm g}$ & 25, 50, 100\\
Cone Start \& End (Geom. \#2) & ($r_1$, $r_2$) & $r_{\rm g}$ & (2.5,20), (10,50), (50,100) \\
Wedge Opening Angle (Geom. \#1)& $\theta_{\rm C}$ & degree & 5,\,45,\,85 \\
Cone Opening Angle (Geom. \#2) & $\theta_{\rm C}$ & degree & 5,\,25,\,45 \\
Corona Temperature & $T_{\rm C}$ & keV & 5, 10, 25, 100, 250, 500\\
Corona Optical Depth & $\tau_{\rm C}$ & none & 0, 0.125, 0.25, 0.5, 0.75, 1, 2 \\
\enddata
\end{deluxetable*}

The parameter grid describing the simulated configurations is summarized in Table \ref{t:conf}. Black holes with spin parameters $a$ between -1 and Thorne's 
theoretical maximum spin $a\,=\,$0.998 \citep{1974ApJ...191..507T} are simulated. 
The sampling is denser close to the maximum spin as the 
accretion disk properties change rapidly as $a$ approaches 1.
We perform the simulations for a black hole with a mass $M$ of 10\,$M_{\odot}$, and 
with different temperature profiles. Taking advantage of the 
analytical result that the temperature profiles 
depend on the accretion rate only 
through a multiplicative scaling factor $\sigma$ \citep{1974ApJ...191..499P}, 
we simulate each model for ten $\sigma$ values.
The simulations with different temperature profiles can 
then be used to derive energy spectra for different 
black hole masses and accretion rates (see Sect. \ref{s:kerrC}).

\begin{figure*}[thb]
\begin{center}
    \includegraphics[width=14cm]{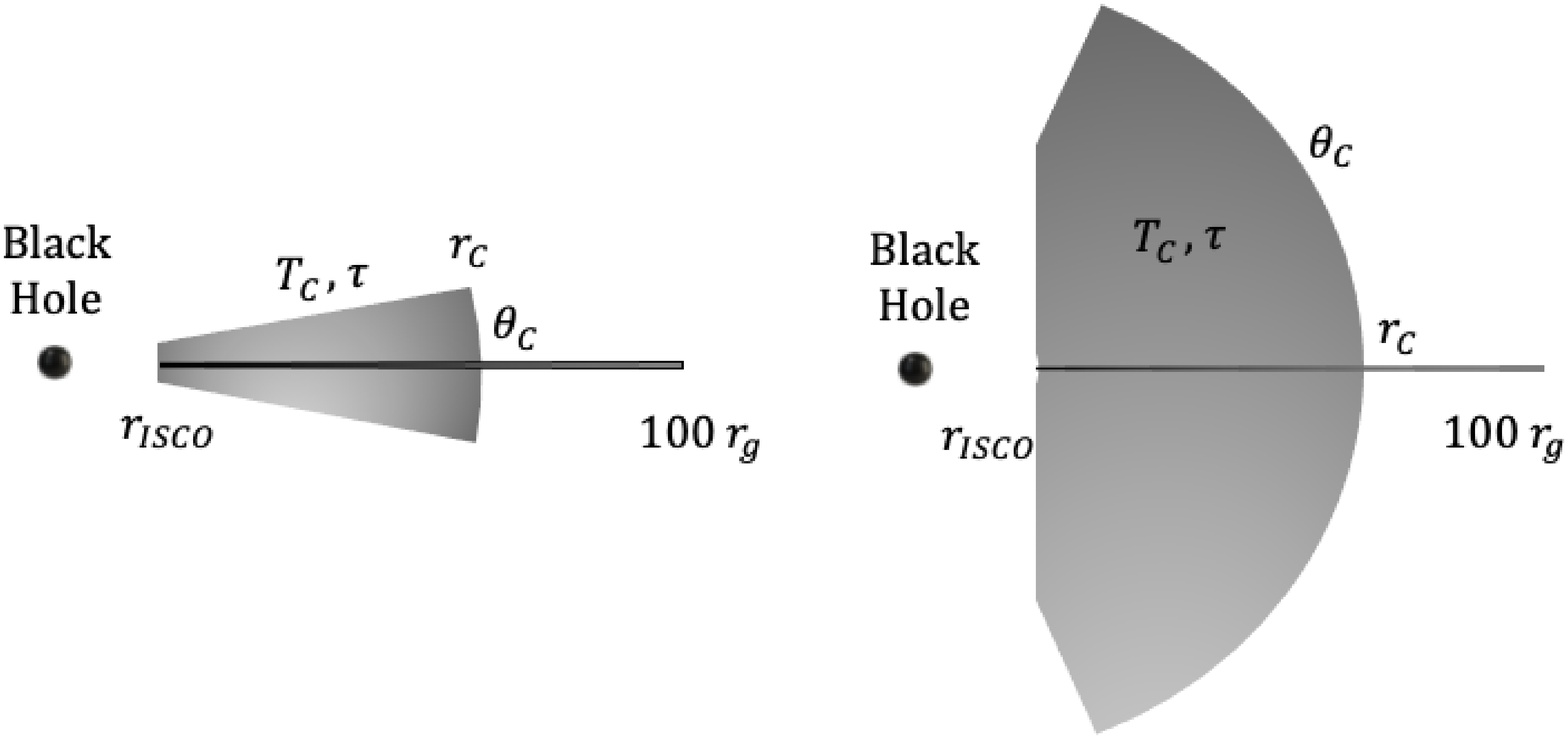}
    \caption{Sketches of simulated wedge-shaped coronae for a small and a large half opening angle $\theta_C$,
    left side and right side, respectively.  
    The accretion disk extends from $r=r_{\rm ISCO}$ to $r=100\,r_{\rm g}$, and the
    corona with electron temperature $T_C$ and vertical optical depth $\tau_C$ extends from
    $r=r_{\rm ISCO}$ to $r=r_C$. 
    \label{f:wedge}}
\end{center}
\end{figure*}

We simulate two families of coronal geometries. 
The first family are wedge-shaped coronae surrounding the accretion disk (Fig. \ref{f:wedge}). 
The coronae extend from the innermost stable circular orbit $r_1=r_{\rm ISCO}$ \citep{1972ApJ...178..347B} 
to $r_2$ equal 25, 50 or 100 $r_{\rm g}$ with a half-opening angle $\theta_{\rm C}$ 
between 5$^{\circ}$ and 85$^{\circ}$ above and below the accretion disk.
Note that the wedge-shaped coronae morph into near spherical coronae as $\theta_{\rm C}$ 
increases from 5$^{\circ}$ to 85$^{\circ}$.

\begin{figure*}[thb]
\begin{center}
    \includegraphics[width=14cm]{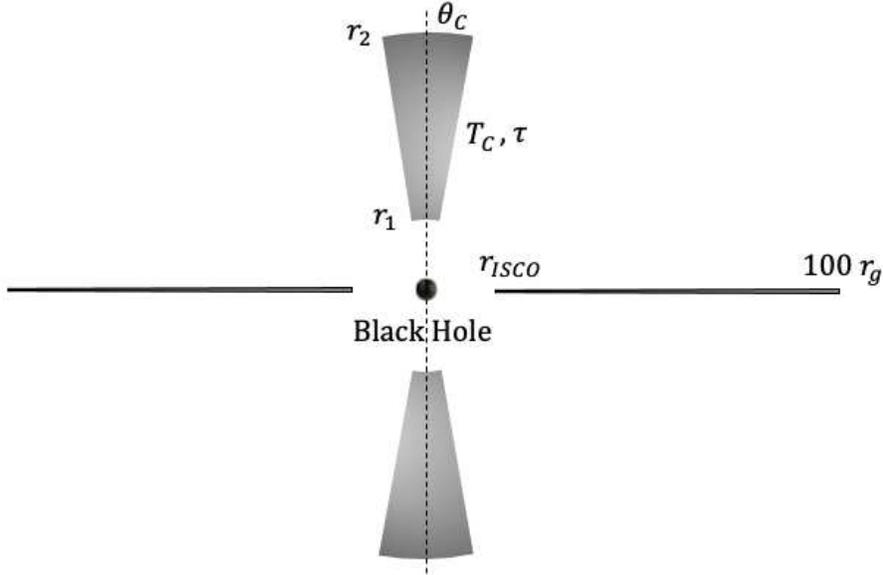}
    \caption{Sketch of simulated cone-shaped corona. The accretion disk extends from $r=r_{\rm ISCO}$ to $r=100\,r_{\rm g}$, and the corona with electron temperature $T_C$ and radial optical depth $\tau_C$ extends from
    $r=r_1$ to $r=r_2$ from the polar angle
    $\theta=0$ to $\theta=\theta_C$. The dashed line shows
    the black hole spin axis. 
        \label{f:cone}}
\end{center}
\end{figure*}

The second family are cone-shaped coronae in the funnel regions around 
the black hole spin axis (Fig. \ref{f:cone}). 
The cone-shaped coronae extend from $r_1\,=\,2.5$\,$r_{\rm g}$, 10\,$r_{\rm g}$, or 50\,$r_{\rm g}$ to
$r_2\,$ 20\,$r_{\rm g}$, 50\,$r_{\rm g}$, or 100\,$r_{\rm g}$, respectively, above and bellow the black hole.
The opening angles $\theta_{\rm C}$ range from 
5$^{\circ}$ to 45$^{\circ}$.
For $\theta_{\rm C}=5^{\circ}$, the cone-shaped coronae resemble a jet in the funnel region above and below the
black hole. For $\theta_{\rm C}=45^{\circ}$, the cone-shaped coronae resemble less collimated structures. 

Corona electron temperatures $T_{\rm C}$ between 5\,keV and 500\,keV, 
and integrated optical depths between the case of no corona ($\tau_{\rm C}\,=\,0$) and
$\tau_{\rm C}\,=\,2$ are simulated.
For wedge-shaped coronae, the optical depth is measured 
vertically from the accretion disk to the upper (or lower) edge of the corona.
For the cone-shaped coronae, the optical depth is measured radially from the inner to the
outer edge of the corona.
For a fast spinning black hole ($a=0.998$)
with a wedge corona extending from the ISCO to $100\,r_{\rm g}$,
the highest temperature ($T_{\rm C}=500$\,keV), 
an optical depth perpendicular to the accretion disk of
($\tau_{\rm C}=2$), and a wedge angle of $20^{\circ}$, 
the code produces a power law component 
with a photon index $\Gamma$ 
(from $dN/dE\propto E^{-\Gamma}$) of $\sim$0.7.

The code tracks photons forward in time, and allows us to collect photons
at arbitrary locations of the observer relative to the accreting system. 
A single simulation set can thus be used to
infer the observations for all black hole inclination angles.
\subsection{Raytracing Code}
We generate the simulated energy spectra with the code {\tt xTrack} 
described in \citep{2012ApJ...754..133K,2016PhRvD..93d4020H,
2017ApJ...850...14B,2019ApJ...870..125K,2020ApJ...889..111A,2021ApJ...906...28A}. 
We give here a slightly updated description of the code that 
includes  several recent improvements.

The code uses the Kerr metric $g_{\mu\nu}$ in Boyer-Lindquist (BL) coordinates 
$x^{\mu}=(t,r,\theta,\phi)$.
Photons are emitted and scatter in the reference frames of the emitting or scattering plasma, 
and are transported in the global BL coordinate frame.  The reference frame transformations 
use orthonormal tetrads.
The code assumes that the black hole accretes through a geometrically paper thin, 
optically thick accretion disk \citep{1973A&A....24..337S,1973blho.conf..343N,1974ApJ...191..499P}. 
For each black hole spin we calculate a fiducial temperature profile $T_0(r)$ with 
the accretion rate $\dot{M}$ chosen such that the luminosity
\begin{equation}
L_{\rm acc}\,=\,\eta(r_{\rm ISCO})\dot{M} \label{e:lacc}
\end{equation}
is 10\% of the Eddington luminosity
\begin{equation}
L_{\rm Edd}\,=\,
\frac{4\pi\,G\,M_{\rm BH}\,m_{\rm P}}{\sigma_{\rm T}}.
\end{equation}
In the above equations, $\eta$ is the fraction of the gravitational energy 
of the matter that is converted to heat as the matter accretes from infinity to 
$r_{\rm ISCO}$ in units of the matter's rest mass energy.
The fraction is given by:
\begin{equation}
\eta(r_{\rm ISCO})\,=\,1-E^{\dagger}(r_{\rm ISCO})
\end{equation}
with $E^{\dagger}(r_{\rm ISCO})$ being the specific energy at infinity 
of the matter at $r\,=\,r_{\rm ISCO}$:
\begin{equation}
E^{\dagger}\,=-\,g_{t\mu}\,u^{\mu}
\end{equation}
where $g$ is the Kerr metric and $u^{\mu}$ the four velocity of the matter at the ISCO.
Given the angular velocity $d\phi/dt=\Omega\,=\pm M^{1/2}/(r^{3/2}\pm a M^{1/2})$ 
\citep{1972ApJ...178..347B}, we get $u^{\mu}\,=\,u^t\,(1,0,0,\Omega)$ from the condition $u^2=-1$.
The emissivity (power per comoving time and area) $F(r)$ is then given by 
Equations\ (11)-(12) of \citep{1974ApJ...191..499P}.
We assume a fiducial temperature profile with 
\begin{equation}
T_0(r)=\left(\frac{F(r)}{\sigma_{\rm SB}}\right)^{1/4},
\label{e:t0}
\end{equation}
where $\sigma_{\rm SB}$ is the Stefan Boltzmann constant.
As mentioned above, we perform the simulations for a range of scaled temperature profiles:
\begin{equation}
    T(r)\,=\,\sigma\,T_0(r)
\end{equation}
with $\sigma$ between 0.05 and 2.
We generate initial photon energies assuming a 
diluted blackbody energy spectrum with 
a hardening factor of $f_{\rm h}=1.8$
\citep{1995ApJ...445..780S}.

For each accretion disk and corona configuration, we simulate 2$\times$10$^7$ photons, 
using 10,000 logarithmically spaced radial bins with $r$ running from 
$r_{\rm ISCO}$ to $100\,r_{\rm g}$. 
The photons are launched into the upper hemisphere with constant
probability per solid angle with a limb brightening weighting factor 
and initial polarization given by Table XXIV of \citep[][]{1960ratr.book.....C}.
We normalize the limb brightening weighting factor so that the 
average weighting factor is 1. 

An adaptive step-size Cash-Karp integrator based on a 5'th 
order Runge-Kutta algorithm is used to solve 
the geodesic equation, to update the the position ($x^{\mu}$), 
and the photon's wavevector $k^{\mu}$, and to parallel transports the 
polarization vector $f^{\mu}$ \citep{2020ApJ...889..111A,2020arXiv200803829A}. 
The polarization fraction remains constant between scattering events.
The step size is reduced if the traversed coronal optical depth evaluated 
in the rest frame of the corona exceeds 2\% of the  total optical depth of the corona.
The step size is furthermore modified if the photon enters or leaves the corona, 
so that the end of the step coincides with the corona boundary.

Photons are tracked until their radial BL coordinate drops below 1.02 times the $r$-coordinate
of the event horizon (at which point we assume that the photon will disappear into the black hole) 
or reach a fiducial observer at $r_{\rm obs}\,=$ 10,000\,$r_{\rm g}$. 
In the latter case, the wave and polarization vectors are transformed into the 
reference system of a coordinate stationary observer, and key information
is written to disk.
\subsection{Calculation of Absolute Fluxes Reaching the Observer and Reaching the Accretion Disk\label{s:sbs}}
Each event is counted with a weighing factor that allows us to make
absolute flux predictions. As mentioned above, the thin disk solution of \citet{1974ApJ...191..499P}
gives the plasma-frame energy flux $F(r)$ of the photons emitted at radius $r$.
The number of photons emitted per Boyer-Lindquist $dt$, $dr$ and $d\phi$ is given by \citep{2012ApJ...754..133K}:
\begin{equation}
\frac{dN}{dt\,dr\,d\phi}(r)\,=\,\frac{\sqrt{-g_{tr\phi}(r)}\,F(r)}{<\!E(r)\!>}
\label{e1}
\end{equation}
with $\sqrt{-g_{tr\phi}(r)}$ is the square root of the negative of $t-r-\phi$-part of the metric.
For the Kerr metric, the factor simplifies to $\sqrt{-g_{tr\phi}(r)}=r$ in the equatorial plane.
The mean plasma frame energy of the thermally emitted photons
is given by 
\begin{equation}
<\!E(r)\!>\,=\,2.7\,f_{\rm h}\,k_{\rm B}\,T_{\rm eff}(r).
\end{equation}
Although Equation (\ref{e1}) has been derived from a relativistic invariant,
the interpretation is simple: 
the photon flux per unit coordinate time $dN/dt$
equals the proper area-time volume element 
of the emitting disk segment $\sqrt{-g_{tr\phi}(r)}\,dr\,d\phi\,dt$ 
times the emitted energy flux per unit area $F(r)$
divided by the product of the mean energy of the 
emitted photons $<\!E(r)\!>$ 
times $dt$.   

The $i$'th disk segment extending from $r_i-\Delta r_i/2$ to $r+\Delta r_i/2$
thus emits photons at a rate of:
\begin{equation}
\frac{dN_i}{dt}\,=\,\Delta r_i\,\int_0^{2\pi}\,d\phi\,\frac{dN}{dt\,dr\,d\phi}(r_i)
\,=\,2\pi \Delta r_i\,\frac{dN}{dt\,dr\,d\phi}(r_i)
\end{equation}
Given that we generate $N_{\rm bin}$ events per radial bin,
each simulated photon transports the photon rate
$\frac{dN_i}{dt}/N_{\rm bin}$.
The photon rate per simulated event launched from the
$i$'th bin is thus give by the weight:
\begin{equation}
w_i=\frac{1}{N_{\rm bin}}\frac{dN_i}{dt}=
\frac{2\pi\,\sqrt{-g_{tr\phi}(r_i)}\,\Delta r_i}{N_{\rm bin}}
\frac{F(r_i)}{<\!E(r_i)\!>}
\label{e2}
\end{equation}

For a source at the distance $d$ from us, the flux spreads over an area:
\begin{equation}
A_{\Omega}\,=\,\Delta \Omega \,d^2
\end{equation}
before reaching us. 
Here, $\Delta \Omega$ is the solid angle of the
$\theta$-window that we use to collect the simulated photons.
Each simulated event thus
transports the photon flux:
\begin{equation}
f_i\,=\,\frac{1}{A_{\Omega}}w_i\,=\,
\frac{2\pi\,\sqrt{-g_{tr\phi}(r_i)}\,\Delta r_i}{N_{\rm bin}\,A_{\Omega}}
\frac{F(r_i)}{<\!E(r_i)\!>}
\label{e:f1}
\end{equation}
per unit observer time and area.
Differential photon fluxes per unit time, unit area, and unit energy (Stokes $I$, $Q$, and $U$)
can be obtained by binning the results in observer energy
and dividing the sum of all $f_i$ in each bin by the 
width of the energy bin. 

The calculation of the reflected energy spectrum
requires the knowledge of the flux and the photon index
of the emission irradiating the $j$'th disk element.
Tracing the argument that led to Equation (\ref{e2}) backwards,
we infer that each event being launched from the $i$'th bin
of width $\Delta r_i$ that reaches the $j$'th bin of 
width $\Delta r_j$ adds a plasma frame energy flux of
\begin{eqnarray}
f_{i\rightarrow j}&=&\frac{E_d\,w_i}{2\pi\sqrt{-g_{tr\phi}(r_j)}\, \Delta r_j} \nonumber \\
&=&\frac{F(r_i)}{N_{\rm bin}}\frac{\sqrt{-g_{tr\phi}(r_i)}\Delta r_i}{\sqrt{-g_{tr\phi}(r_j)}\Delta r_j} \frac{E_d}{<E_i>}
\label{e:f2}
\end{eqnarray}
to the energy flux irradiating the bin.
Here, $E_d$ is the plasma frame energy of the photon irradiating the disk. 
This equation has again a simple interpretation: the energy flux transported by each event 
starting at bin $i$ and reaching bin $j$ is the energy flux $F(r_i)$ leaving bin\,$i$ divided by the
number of events launched from the $i$'th bin times 
the ratio of the area-time volumes of 
the $i$'th and $j$'th radial bins times the fractional energy gain or loss
of the photon between emission from bin $i$ and arrival at bin~$j$.

Summing $f_{i\rightarrow j}$ over all events irradiating the $j$'th radial bin gives the total 
plasma frame energy flux $F_x(r_j)$ irradiating the bin.
Reflections off the disk and/or Compton scattering 
processes modify the statistical weight of a photon.

In a photospheric plasma with comoving electron density $n_{\rm e}$
\citep{1993MNRAS.261...74R,1996MNRAS.278.1082R,1996MNRAS.281..637R,1999MNRAS.306..461R}, 
the ionization can be characterized with the ionization parameter $\xi$.
The parameter is proportional to the ratio of the 
photoionization rate ($\propto n_{\rm e}$) divided by the 
recombination rate ($\propto n_{\rm e}^{\,2}$):
\begin{equation}
    \xi(r_j)=\frac{4\pi F_x(r_j)}{n_{\rm e}(r_j)}.
    \label{e:xi}
\end{equation}
The energy spectra of all photons arriving in the radial bin $j$ can be used to
fit the photon index $\Gamma_j$ (from $dN/dE\propto E^{-\Gamma}$)
of the emission irradiating the bin. 

A limitation of our code should be mentioned: 
photons reflecting off the disk more than once are not
modeled fully self-consistently. For these photons, only their 
first encounter with the disk contributes to the calculation 
of $\xi$ and $\Gamma$, and additional encounters are neglected.
The statistical weight of the subsequent 
photon-disk encounters depends on the $\xi$ and $\Gamma$ values used 
for the previous encounters. Properly accounting for multiple 
disk-photon interactions would thus render the problem non-linear.
For the purpose of calculating the predicted energy spectra, {\tt kerrC} 
does track the photons through multiple disk encounters.
However, the {\tt XILLVER} tables are only used to modify their 
statistical weight for their first disk encounter. 
For subsequent encounters, the weights from the Chandrasekhar 
treatment are used.
We estimated the impact of these approximations on the fitted parameters, 
by using the ``opposite'' assumptions, i.e. by including the additional photon-disk encounters in the $\xi$ and $\Gamma$ calculation with their full
Chandrasekhar weight, and by completely excluding photons scattering 
more than once from entering the predicted energy spectra.
We find that the approximations have a negligible impact 
in case of the cone-shaped corona models as few photons 
return multiple times to the disk. Adopting the alternative 
assumptions for the wedge-shaped coronae tends to harden 
the energy spectra impinging on the disk for the 
high-$\xi$ models and to soften the predicted energy spectra 
for the low-$\xi$ models.
We will explore the problem and possible solutions in greater detail in future papers.
\subsection{Comptonization}
We use the Comptonization code of \citep{2017ApJ...850...14B,2018PhDT........98B}.
The coronal plasma is assumed to be stationary in the Zero Angular Momentum Observer (ZAMO) 
frame \citep{1972ApJ...178..347B}. See \citep{2021ApJ...906...34K} for a formalism to 
implement relativistically moving coronal gas. 
For each integration step, the traversed Thomson optical depth $\Delta \tau$ 
is calculated and a random number generator decides if the photon is considered 
for a coronal scattering event. The scattering event is simulated by 
transforming the photon's wave and polarization four vectors into the ZAMO frame, 
and subsequently into the frame of a scattering electron. 
We assume that electrons move isotropically in the ZAMO frame with energies drawn from a 
relativistic Maxwell–Jüttner distribution \citep{1911AnP...339..856J}.
The scattered photon direction is drawn from a distribution with equal 
scattering  probability in the  electron rest frame. 
Subsequently, the Stokes parameters are calculated relative to the scattering plane. 
The fully relativistic Fano scattering matrix derived from the Klein-Nishina 
cross section \citep{1949JOSA...39..859F,RevModPhys.33.8} 
is used to calculate the Stokes vector of the outgoing beam. 
We use the Stokes $I$ component of the scattered beam and a random number generator
to decide if a photon actually scatters. This rejection technique enables us to account 
for the dependence of the scattering probability on the scattering direction and on the photon energy.
If the photon indeed scatters, the statistical weight is multiplied with 
the kinematic factor $1-\cos{(\theta)}$ with $\theta$ being the pitch 
angle of the incoming photon and electron, accounting for the relative 
motion of the photons and electrons \citep[][]{2017ApJ...850...14B}. 
The scattering is followed 
by the back transformation of the photon wave and polarization vectors into the BL frame.
\subsection{Preliminary Disk Reflection\label{s:chandra}}
For photons irradiating the accretion disk, we perform a
simple reflection in the reference frame of the disk plasma
based on the analytical results of Chandrasekhar for the reflection of a polarized beam of photons off an indefinitely deep plane-parallel atmosphere \citep[][Section 70.3, Equation (164) and Table XXV]{1960ratr.book.....C}.
The reflection changes the statistical weight of the reflected beam, and the linear polarization fraction and angle. 
For each reflected photon, the code saves
information about the incident photon and 
the reflected photon as measured in the
disk reference frame, and information about
the photon when it reaches the observer.
This information is used during the actual fitting 
of the observational data to re-weigh the reflected 
photons, so that the reflected energy spectrum resembles 
that from the {\tt XILLVER} radiation transport 
calculations for the self-consistently derived 
ionization parameter $\xi(r)$ and the  
spectral index $\Gamma(r)$ 
of the emission irradiating
the accretion disk.\\[2ex]

\section{Implementation of the {\tt kerrC} Model}
\label{s:kerrC}
\subsection{Implementation Details}
The {\tt kerrC} fitting model uses a data bank of
1.36 trillion simulated events 
(20 million events for each of the 
\config~configurations.)
The raytracing code stores for this purpose 
three types of data: data of photons reaching the observer without reflecting off the disk
(direct emission data), data of photons irradiating the disk (disk data), 
and data of photons reaching the observer after scattering at least once off the
accretion disk (reflected emission data). 
This classification is independent of the scattering of photons in the corona.
The data are stored in Hierarchical Data Format version 5 (HDF5) files\footnote{https://portal.hdfgroup.org/} which makes it possible to quickly access the information of the
photons that matter for the considered region 
of the parameter space. 

\begin{deluxetable*}{|c|p{5cm}|c|c|c|p{2.5cm}|}[t!]
\tablenum{2}
\tablecaption{{\tt kerrc} model parameters \label{t:par}}
\tablewidth{0pt}
\tablehead{
\colhead{Number} &\colhead{Parameter} & \colhead{Symbol} & \colhead{\kerrc name} & \colhead{Unit} & \colhead{Allowed Values}
}
\startdata
\multicolumn{6}{|c|}{Black Hole Parameters}\\ \hline
1 & Black Hole Mass & $M$ & {\tt M} & $M_{\odot}$ & any \\
2 &Black Hole Spin Parameter & $a$ & {\tt a} & none & -1 ... 0.998 \\
3 & Black Hole Inclination & $i$ & {\tt incl} & degree & 5 ... 85 \\
4 & Distance & d & {\tt dist} & kpc & 0 ... $\infty$ \\ \hline
\multicolumn{6}{|c|}{Accretion Flow Parameters}\\ \hline
5 & Mass Accretion Rate & $\dot{M}$ & {\tt mDot} & $10^{18} \,{\rm g\, s}^{-1}$ & depend on $M$ \\ \hline
\multicolumn{6}{|c|}{Corona Parameters}\\ \hline
6 & Corona Geometry & $g$ & {\tt geom} & none & 0, 1 \\
7 &Corona Edge$^{a}$ & $r_{\rm C}$ & {\tt rC} & $r_{\rm g}$ & 2.5 ... 100 \\
8&Corona Opening Angle & $\theta_{\rm C}$ & {\tt thetaC}& degree & 5 ... 85 \\
9&Corona Temperature & $kT_{\rm C}$ & {\tt tempC} & keV & 5 ... 100 \\
10&Corona Optical Depth & $\tau_{\rm C}$ & {\tt tauC} & none & 0 ... 2 \\ \hline
\multicolumn{6}{|c|}{Reflection Parameters}\\ \hline
11&Reflection Amplitude$^b$ & - & {\tt refl} & none & 0 ... 10 \\ 
12&Metallicity & $A_{Fe}$ &  {\tt AFe} &  solar & 0.5-10\\
13&Electron Temp. & $kT_{\rm e}$ & {\tt kTe}&keV & 1-400 \\
14&Electron Density at $r_{\rm ISCO}$& log$_{10}\,n_{\rm e,0}$ & {\tt logDens}  & cgs & 15-22\\ 
15&Radial Density Powerlaw Index & $\alpha$ & {\tt xiIndex} & none & 0 ...  \\ 
16&$\Gamma$-fit range$^{c}$ & $\kappa$ & {\tt kappa} & none & 0.1...10 \\ \hline 
\multicolumn{6}{|c|}{Polarization Parameters}\\ \hline
16&Angle of BH Spin Axes from Cel. North Pole, counter-clockwise & $\chi$ & {\tt chi} & degree & 0 ... 180 \\ \hline
\multicolumn{6}{|c|}{Additional Model Modifiers}\\ \hline
17&Inner Cutoff$^{d}$  & $l_1$ & {\tt l1} & $r_{\rm g}$ & 0, $r_{\rm ISCO}$ ... 100 \\
18&Outer Cutoff$^{d}$  & $l_2$ & {\tt l2} & $r_{\rm g}$ & 0, $r_{\rm ISCO}$ ... 100 \\
\enddata
$^{a}$ The corona edge parameter gives the outer edge of the wedge-shaped corona (25 ... 100 $r_{\rm g}$) and the inner edge of the cone-shaped corona (2.5 ... 50 $r_{\rm g}$).\\
$^{b}$ The reflection amplitude should be set to unity to get the self-consistently calculated reflection.\\
$^{c}$ {\tt XILLVER} requires the photon indices of the emission irradiating the accretion disk.
We fit the energy spectra from $\kappa E_1$ to $\kappa E_2$ with $\kappa\,=\,1.5$, $E_1\,=\,1$\,keV and
$E_2\,=\,10$\,keV.\\
$^{d}$ $l_1$ ($l_2$) give a lower (upper) bound on the radial coordinate of the emission of photons entering the analysis. 
\end{deluxetable*}

The fitting code is implemented as a user model for the
{\tt Sherpa} fitting package \citep{2001SPIE.4477...76F,2007ASPC..376..543D,SciPyProceedings_51,brefsdal-proc-scipy-2011}. 
While the user model uses a {\tt Python} interface, the work of reading and convolving the raytraced photon and the reflection model data, and interpolating between the
simulated grid points is done by a fast
C++ code compiled into a Python module
with the help of the Boost-Python library 
(D.\ Abrahams and R.\ W.\ Grosse-Kunstleve)
of the Boost distribution\footnote{https://www.boost.org}.

\subsection{Fitting Parameters}
The \kerrc model parameter are listed in Table \ref{t:par}. 
The  parameters include the black hole mass $M$,
the black hole spin $a$, black hole inclination $i$, and
distance $d$. The accretion flow is characterized 
by a single parameter, the mass accretion rate $\dot{M}$. 
The model parameter $g$ selects between the wedge-shaped 
and cone-shaped corona configurations characterized by 
$r_C$, $\theta_C$, $T_C$, and $\tau_C$ as described above.

The reflection parameters include an amplitude to increase the reflected emission above or below
the self-consistently derived intensity. 
The other reflection parameters describe the physical conditions in the photosphere of the
accretion disk. The parameters include the metallicity $A_{Fe}$ relative to solar,
the electron density $n_{\rm e}$ at the inner edge of the accretion disk,
and a power law index giving how the electron density scales with radial distance 
($n(r)\propto r^{-\alpha}$). 
{\tt XILLVER} assumes that the photons irradiating the accretion disk have an energy spectrum of disk photons comptonized by electrons of temperature $kT_{\rm e}$ according to the {\tt Nthcomp} model \citep{1996MNRAS.283..193Z,1999MNRAS.309..561Z}. 
We consider $kT_{\rm e}$ as a free fitting parameter here. 
The parameter $\kappa$ modifies the energy range used for determining the photon index of the
emission irradiating the disk. We use $\kappa=1.5$ giving a fitting range from 1.5 keV to 15 keV.
As mentioned above, the code calculates all three Stokes parameters.
The model parameter $\chi$ gives the rotation 
of the black hole spin axis of the model 
relative to the celestial north pole 
(the black hole spin axis of the model 
is turned counter-clockwise for $\chi>0$). 
Finally, the user can exclude photons launched at radial distances
outside the $l_1$ ... $l_2$ interval from the analysis.
This can be used to estimate the impact of 
disk truncation or shadowing in rough approximation.

\subsection{Model Evaluation}
As described above, we simulated a library for a set of black hole, accretion disk, and
corona configurations. When {\tt kerrC} is called with parameters 
between the simulated ones, {\tt kerrC} identifies the nearby simulated configuration nodes, 
calculates energy spectra for these nodes, and returns an
interpolated energy spectrum using linear interpolation in simplices \citep{weis:89}.

For each configuration node, the direct emission data are used
to calculate the energy spectrum of the direct emission.
Subsequently, the disk data are analyzed to obtain the 
radial 0.1-1000 keV fluxes $f_i$
and the 1.5-15 keV photon indices $\Gamma_i$ 
of the returning and corona photons irradiating the disk 
in logarithmicly spaced radial bins ($i=1...12$).
For the same radial bins, the plasma frame 
energy spectra $F_{C,i,j}(E)$ of photons scattered into 
each of 10 inclination bins ($j=1...10$) are acquired. 
The inclination refers to the plasma frame direction 
of the scattered photon.  
The subscript $C$ indicates that these energy
spectra are based on Chandrasekhar’s scattering formalism. 

In the last step, the reflected energy spectra are calculated.
For this purpose, each reflected photon enters the analysis
with a weight that accounts for the results from the {\tt XILLVER} 
radiative transport calculations. 
We use the energy spectra 
$F_X(\Gamma,A_{Fe},\log_{10}\xi,kT_{\rm e},\log_{10}n_{\rm e},\theta;E)$
from the {\tt XILLVER} tables version Cp\_3.6 
\footnote{http://www.srl.caltech.edu/personnel/javier/xillver/}.
Here, $\Gamma$ is the photon index of the emission irradiating the accretion disk, 
$A_{Fe}$ is the metallicity relative to solar system metallicities, 
$\xi$ is the ionization parameter, $kT_{\rm e}$ is the electron temperature
describing the Comptonized  {\tt XILLVER} input energy spectrum,
$n_{\rm e}$ is the electron density of the reflecting plasma, $\theta$ is the
plasma frame inclination angle of the reflected emission, 
and $E$ is the plasma frame energy of the incoming and outgoing photon. 

Our treatment seeks to minimize the impact of the {\tt XILLVER} input 
energy spectra by multiplying each photon’s statistical weight with
the ratio $r$ of the energy spectrum for the actual ionization
parameter divided by the energy spectrum 
for the highest simulated 
ionization parameter ($\log_{10}\xi_{\rm max}=4.7$):
\begin{equation}
    r\,=\,\frac
    {F_X(\Gamma_i,A_{Fe},\xi(r_i),kT_{\rm e},
    n_{\rm e}(r_i),\theta;E)}
    {F_X(\Gamma_i,A_{Fe},\xi_{\rm max},kT_{\rm e},
    n_{\rm e}(r_i),\theta;E)}
    \label{e:r}
\end{equation}
The index $i$ denotes the radial bin $r_i$ where the reflection takes place,  
$\Gamma_i$ and $\xi(r_i)$ are known from the disk analysis, 
$kT_{\rm e}$ is a fit parameter, $n_{\rm e}(r_i)$ follows from the fit parameters $n_{\rm e,0}$ and $\alpha$, and
$\theta$ and $E$  are the inclination and energy of the reflected photon. 
The multiplicative correction is justified by the fact that the {\tt XILLVER} results
for $(\log_{10}\xi)_{\rm max}=4.7$ agree with Chandrasekhar’s results for a
pure electron scattering atmosphere
\citep[][and private communication]{2010PhDT.......139G}.
The motivation for our treatment is that the reflected Chandrasekhar energy spectrum resembles the reflected energy spectrum to first approximation. 
The re-weighting factor is used 
as a correction of the Chandrasekhar result. The treatment leads to more physical results than using the ratio of the {\tt XILLVER} energy spectrum 
divided by the incident energy spectrum.
In the latter case, the {\tt XILLVER} assumption of incident power law
energy spectra can lead to reflected 
energy spectra with unphysically 
high fluxes of high-energy photons, exceeding by far the high-energy photon fluxes provided by the corona.
The {\tt XILLVER} tables are binned in 2999 logarithmicly spaced energy bins from 0.07 to 1000.1 keV (adjacent bins spaced 0.3\% apart). We smooth the {\tt XILLVER} results with a Gaussian kernel with a $\sigma$ of 6 bins (1.9\%) to reduce statistical errors associated with a poor sampling of the {\tt XILLVER} tables  by the simulated photons.

Although our reflection treatment conserves the plasma frame photon energy for 
each individual photon, the reweighing with the {\tt XILLVER} results
accounts for energy gains and losses when averaged over many photons. 
The code uses linear interpolation in simplices 
to get the {\tt XILLVER} energy spectra 
between the simulated {\tt XILLVER} nodes. 
If we hit the edge of the simulated parameter space, we use the edge value. 

Our code uses Chandrasekhar's results for the polarization of photons
scattering off an indefinitely deep electron atmosphere.
For each reflection of a photon, the result accounts for
the polarization fraction and direction and the  
inclination and azimuth angles of the incoming and outgoing
photon beams.
The polarization treatment is thus rather detailed, 
but neglects the impact of atomic emission and absorption.
As scattering tends to generate polarization, but 
atomic emission does not, our predicted Fe\,K-$\alpha$ polarizations
are expected to be slightly too high.
The treatment furthermore does not account for the difference of the
polarization dependence between the Thomson and Klein-Nishina 
scattering cross sections.

Some {\tt XILLVER} models did not converge satisfactorily 
(J. Garc\'ia, private communication).
For electron densities $n_{\rm e}>10^{19}$ cm$^{-3}$ we 
thus use the {\tt XILLVER} model for $n_{\rm e}=10^{19}\,\rm cm^{-3}$.

The code saves CPU time by storing and later re-using 
some of the intermediate results. For example, 
as long as $l_1$ and $l_2$ are not changed, 
the direct emission results and the disk analysis results 
can be stored and re-used for each configuration node.
The reflected emission results need to be re-calculated 
every time the reflection parameters are changed.

\begin{figure*}[t!]
\gridline{\fig{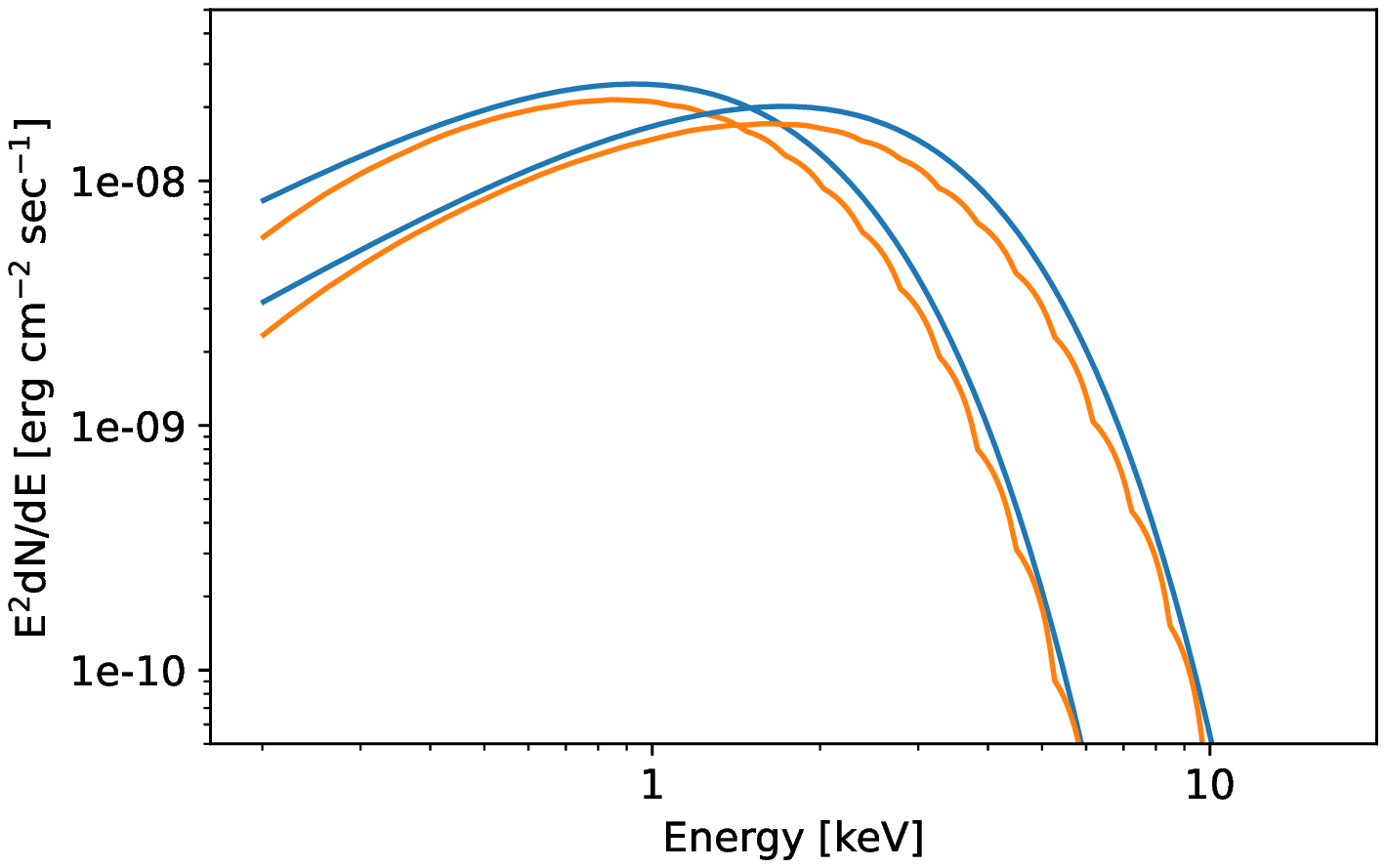}{0.45\textwidth}{(a)}
          \fig{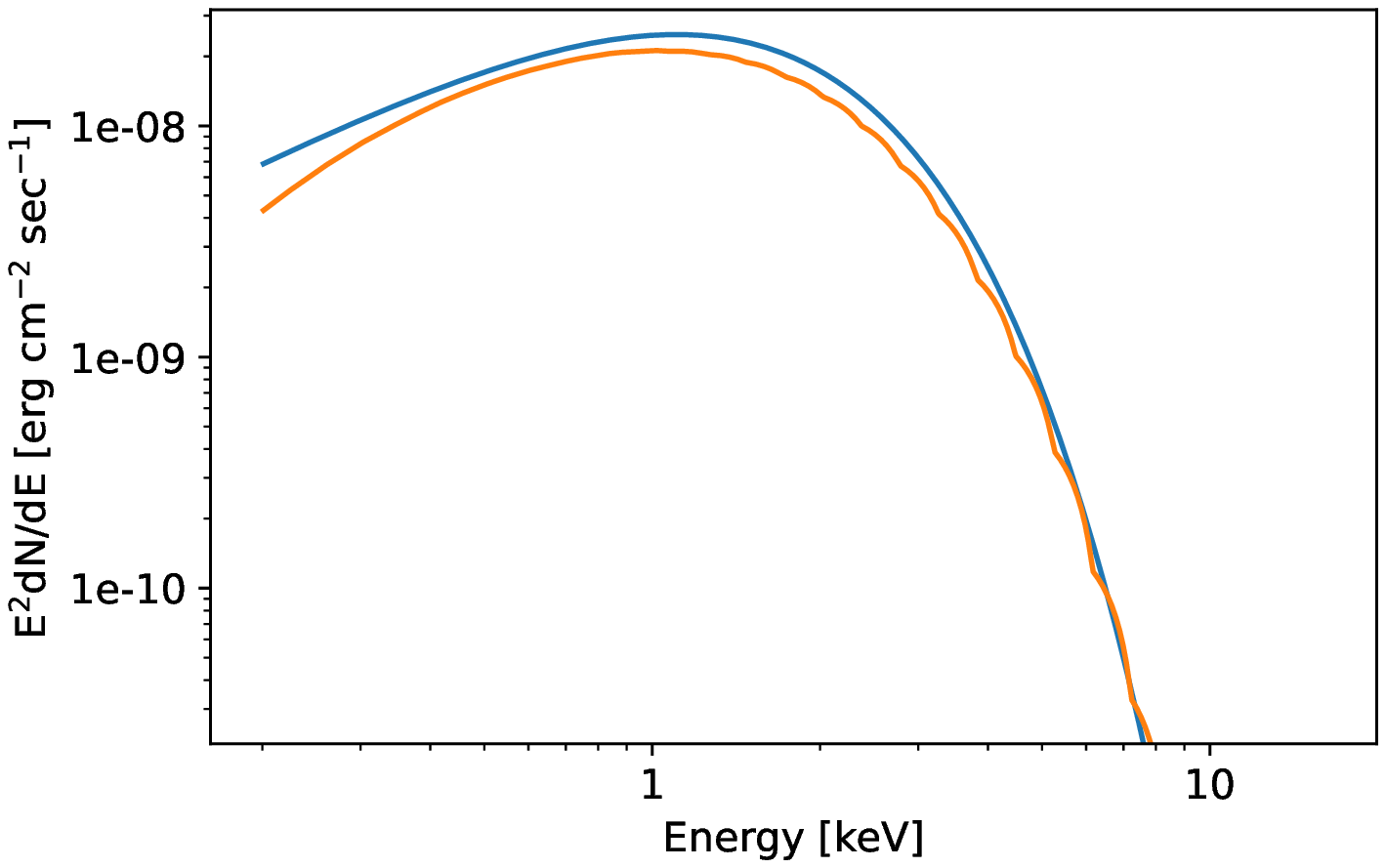}{0.45\textwidth}{(b)}}
\gridline{\fig{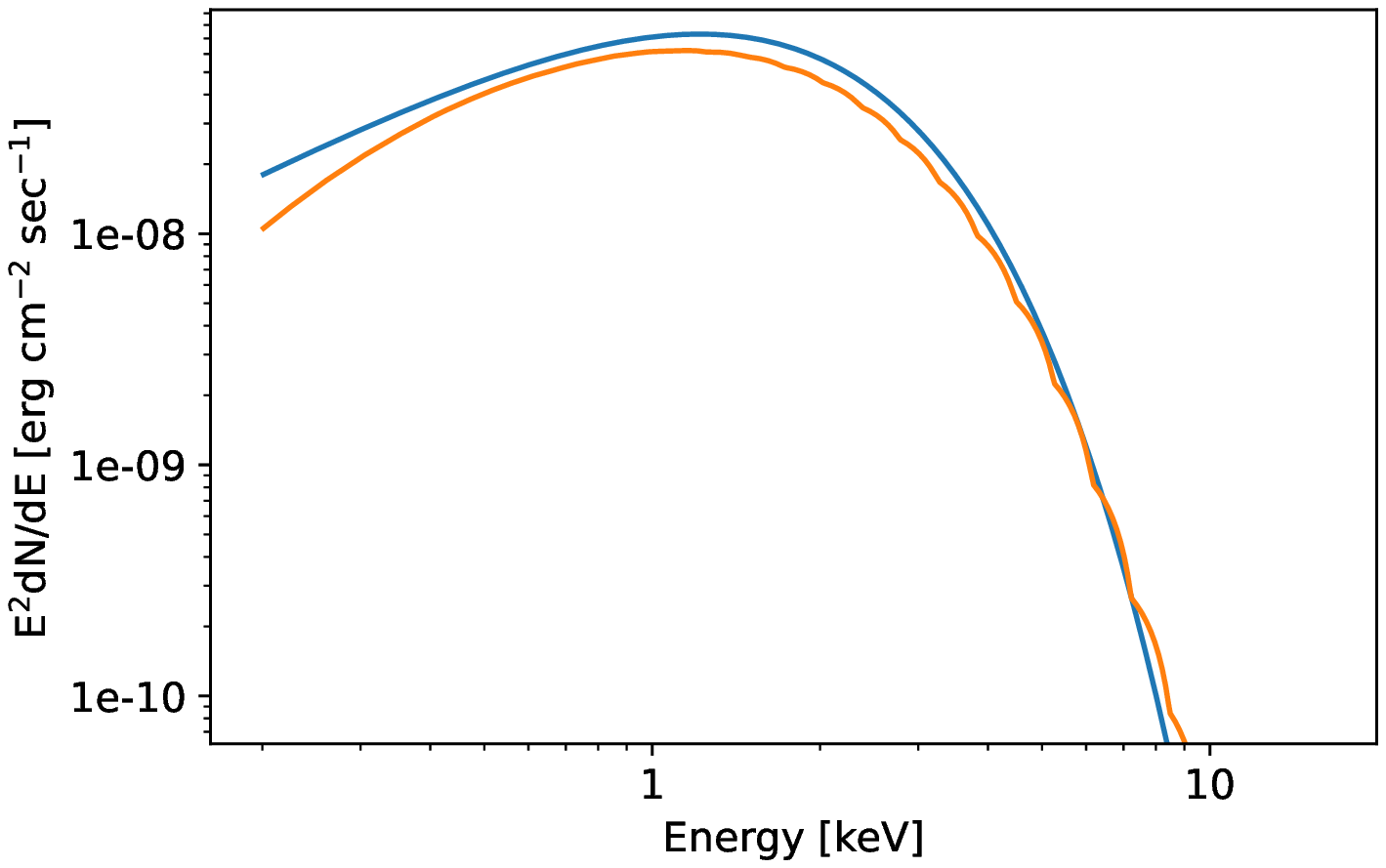}{0.45\textwidth}{(c)}
          \fig{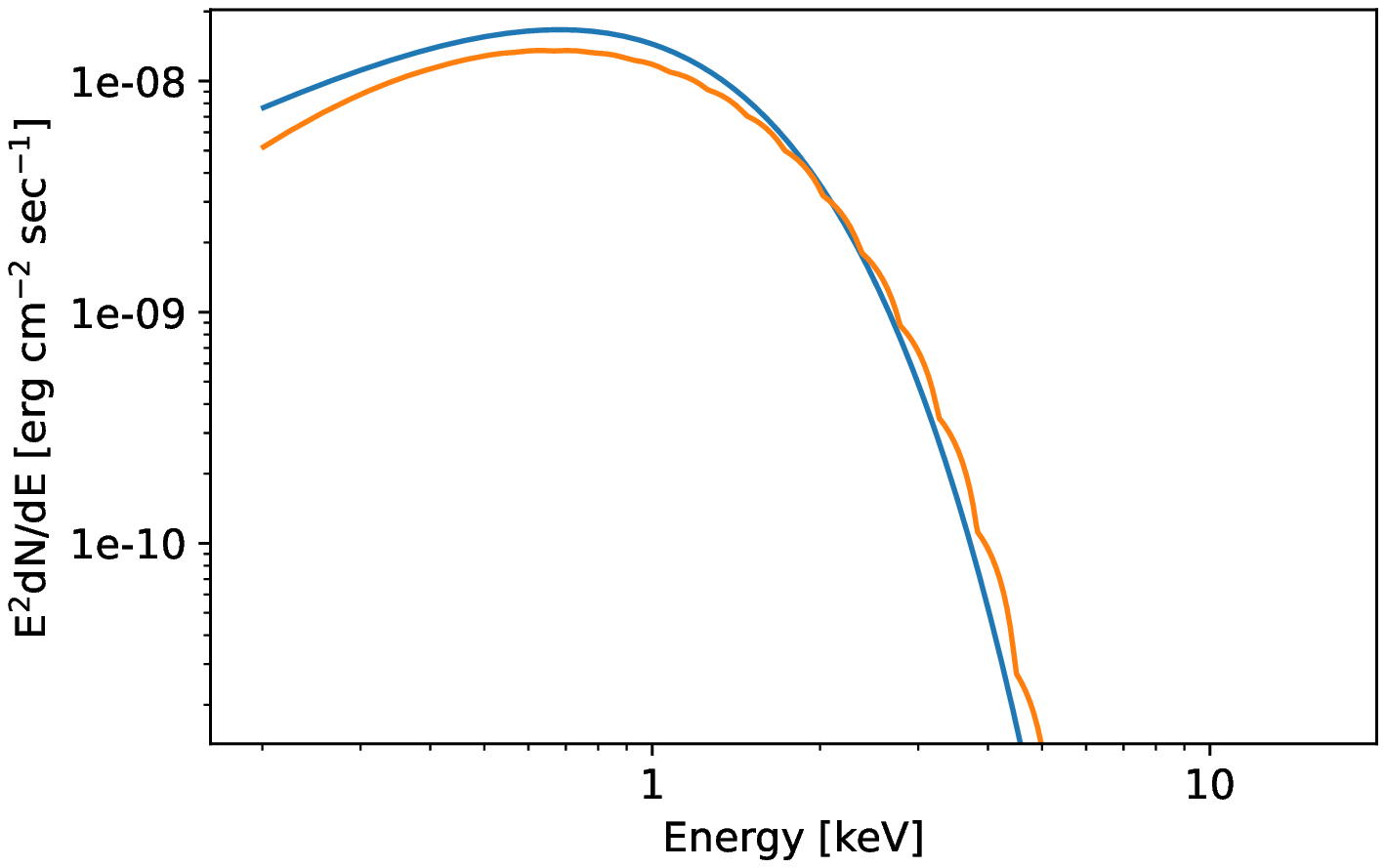}{0.45\textwidth}{(d)}}
\caption{Comparison of {\tt kerrC} (orange lines) and {\tt KERRBB} (blue lines) \citep{2005ApJS..157..335L}. 
Panel (a) shows the models for a spin $a$ of 0.98, inclinations $i$ of $27.51^{\circ}$ (left)
and $65^{\circ}$ (right), a black hole mass of 21.2\,$M_{\odot}$,
an accretion rate $\dot{M}$ of 0.17$\times 10^{18}$\,g/sec, a source distance of 2.22\,kpc, a hardening factor of 1.8, reflected returning emission switched off, and with limb brightening.
The other three panels compare the two models for the
$i\,=\,27.51^{\circ}$ case of (a) when one of the 
parameters is changed but all others are held constant, 
i.e. when the black hole mass is decreased to 15\,$M_{\odot}$
(b), when $\dot{M}$ is increased to 0.5$\times 10^{18}$\,g/sec (c),
and when the spin parameter $a$ is reduced to 0.75 (d). \label{f:comp}}
\end{figure*}

\subsection{Scaling Relations}
The thin disk solution of \citet{1974ApJ...191..499P} 
implies the following scaling relations 
of the temperature scale $\sigma$, 
the observer flux multiplier $f_i$,
and the disk flux multiplier $f_{i\rightarrow j}$
with black hole mass $M$, mass accretion rate
$\dot{M}$, and source distance $d$:
\begin{eqnarray}
\sigma&\propto&M^{-1/2}\,\dot{M}^{1/4} \label{e:s1}\\
f_i&\propto&M^{1/2}\,\dot{M}^{3/4}\,d^{-2} \label{e:s2}\\ 
f_{i\rightarrow j} &\propto &M^{-3/2}\,\dot{M}^{3/4}. \label{e:s3}
\end{eqnarray}
We extracted these scaling relations by numerically calculating
the temperatures and weighting factors for different
$M$ and $\dot{M}$ input parameters, and evaluating 
how they depend on these input parameters.
{\tt kerrC} uses Equation (\ref{e:s1}) to map the 
$M$ and $\dot{M}$ values provided by the user to 
a temperature scale $\sigma$, and then uses this
$\sigma$ to interpolate between the simulated $\sigma$-values.
Equations (\ref{e:s2}) and (\ref{e:s3}) are used to adjust the 
flux multipliers in the analyses of the direct emission,
the emission irradiating the disk, and the reflected emission.

\subsection{Code Validation} 
We validate \kerrc by comparing the predictions for the thermal energy spectrum with the results of {\tt KERRBB} 
\citep{2005ApJS..157..335L}.
Figure \ref{f:comp} shows that the comparison of the 
{\tt kerrC} and {\tt KERRBB} energy spectra typically agree within $<\sim$20\%. 
The most pronounced differences can be recognized at energies below 0.5\,keV.
The fact that {\tt kerrC} underpredicts those fluxes can be explained by
the fact that it only accounts for the emission from $r\le100\,r_{\rm g}$.
At larger radial distances, the accretion disk emits at rather low energies.
Some additional differences may stem from {\tt KERRBB} using a 
limb brightening proportional to $1+\cos{(\theta)}$ and {\tt kerrC} using 
the limb brightening prescription from Chandrasekhar's indefinitely 
deep electron scattering atmosphere. Overall, the energy spectra 
of the two codes agree well.

We have validated the flux energy spectrum and polarization energy spectrum of 
{\tt kerrC} with that of the code {\tt MONK} \citep[][]{2019ApJ...875..148Z} 
and found excellent agreement (Zhang et al., private communication).

\section{Fit of intermediate-state Cyg\,X-1 data with {\tt kerrC}}
\label{s:res}
We show here results from using {\tt kerrC} to fit the 
{\it NuSTAR} \citep{2013ApJ...770..103H} 
and {\it Suzaku} \citep{2007PASJ...59S...1M} observations of the 
archetypical black hole Cyg\,X-1 in the intermediate state.
The data were acquired on May 27-28, 2015 and include 
19,860 sec and 20,500 sec of {\it NuSTAR} 
Focal Plane Module A and B observations, respectively.
We analyzed the data with the {\tt NuSTARDAS} analysis pipeline
provided with the software package {\tt HEAsoft 6.28}.
We use furthermore 4,991 sec of {\it Suzaku} data
aquired with the X-ray Imaging Spectrometer \#1 \citep[XIS1][]{2007PASJ...59S..23K} on the same days.
The {\it NuSTAR} and {\it Suzaku} data have been 
published in \citep{2018ApJ...855....3T}.
The authors kindly shared the {\it Suzaku} XIS1 data with us.
We only use the 3-20 keV {\it NuSTAR} data, as the statistical errors of
the {\tt kerrC} model become appreciable above this energy.
Following \citet{2018ApJ...855....3T}, we use only the
1-1.7 keV and 2.1-8 keV {\it Suzaku} XIS1 data.  

\begin{deluxetable}{ccp{2cm}}
\tablenum{3}
\tablecaption{Best-fit wedge-shaped corona model. \label{t:bf0}}
\tablewidth{0pt}
\tablehead{
\colhead{Parameter} & \colhead{thawed?} & 
\colhead{Result (90\% CL uncertainty)}}
\startdata
\multicolumn{3}{c}{Absorption}\\ \hline
$n_{\rm H}$ & frozen &0.6$\times10^{22}$\,cm$^{-2}$ \\ \hline
\multicolumn{3}{c}{Black Hole Parameters}\\ \hline
$M$ & frozen & 21.2 $M_{\odot}$ \\
$a$ & thawed & \wa \\
$i$ & frozen & 27.51$^{\circ}$ \\
$d$ & frozen & 2.22\,kpc\\ \hline
\multicolumn{3}{c}{Accretion Flow Parameters}\\ \hline
$\dot{M}$ & thawed & \wMdot $\times 10^{18} \,{\rm g\, s}^{-1}$ \\
$r_{\rm C}$ & thawed & \wrCa\,$r_{\rm g}$ \\
$\theta_{\rm C}$ & thawed & \wthetaC$^{\circ}$\\
$kT_{\rm C}$ & thawed & \wtempC\,keV \\
$\tau_{\rm C}$ & thawed & \wtauC \\ \hline
\multicolumn{3}{c}{Reflection Parameters}\\ \hline
$A_{Fe}$ &  frozen & 1 solar\\
$kT_{\rm e}$ & thawed & \wkte\,keV \\
$\log_{10}n_{e,0}$ & thawed  & \wlogNe \\ \hline
\multicolumn{3}{c}{Other Parameters}\\ \hline
$\alpha$ & frozen & 0 \\ 
$l_1$ & frozen & $r_{\rm ISCO}$ \\
$l_2$ & frozen & $100\,r_{\rm g}$ \\ \hline
\multicolumn{3}{c}{Fit Statistics}\\ \hline
$\chi^2/$DoF & NA & \wchi/\dof \\
\enddata
\end{deluxetable}

The estimates of the distance and mass of Cyg\,X-1 have recently been
revised \citet{2021Sci...371.1046M}. 
The new distance estimate accounts for the impact of an
orbital phase dependent attenuation of the radio signal, and 
localizes the source at a distance of $d\,=$\,(2.22+0.18-0.17)\,kpc.
The revised distance and optical data constrain the black hole mass
to be $M\,=$\,21.2$\pm$2.2 $M_{\odot}$. The orbital plane inclination is $27.51^{\circ}$ with 68\% confidence level lower and upper bounds of 
26.94$^{\circ}$ and 28.28$^{\circ}$.

We fit the {\tt kerrC} model absorbed with a fixed $n_H=0.6\times10^{21}$ \citep{2018ApJ...855....3T}
with the cross sections of \citet{1996ApJ...465..487V} and the elemental 
abundances of \citet{2000ApJ...542..914W}.
We freeze $M$, $d$, $i$, and $n_{\rm H}$, 
and assume an accretion disk plasma with solar
elemental abundances ($A_{Fe}\,=\,1$). 

We fit the {\tt kerrC} parameters 
$\dot{M}$, $r_{\rm C}$, 
$\theta_{\rm C}$, $kT_{\rm C}$, $\tau_{\rm C}$ and $\log_{10}n_{\rm e}$.
We assume that the electron density of the photosphere does not depend 
on the radial coordinate ($\alpha\,=\,0$), and we do
not investigate the impact of a truncated 
or shadowed disk. We fitted the data by first 
evaluating the model at all simulated
configuration nodes for one particular set of  
{\tt XILLVER} parameters.
Subsequently, we started the fitting with {\tt Sherpa} 
from the configuration node with the smallest $\chi^2$-value. 
We got the smallest $\chi^2$-values by randomly thawing 
four parameters at a time and repeating the fit, 
switching between the {\tt levmar} and {\tt moncar}
minimization engines. 
We ran the minimization many times with different starting values.
A good number of fits gets stuck in local minima.

A wedge-shaped corona gives the best fit with a $\chi^2$ 
of \wchi\,for \dof\,degrees of freedom (DoF).
The best cone-shaped corona gives a $\chi^2/$DoF of \fchi$/$\dof.
In the following, we discuss the results for the two corona models in turn.
The $\chi^2$/DoF-values are significantly larger than unity showing that 
systematic rather than statistical errors dominate the error budget. 
We thus refrain from giving statistical error estimates.
A detailed error analysis is outside of the scope of this paper.
\subsection{The Best-Fit Wedge-Shaped Corona Model}
\begin{figure}[t!]
\includegraphics[width=0.5\textwidth]{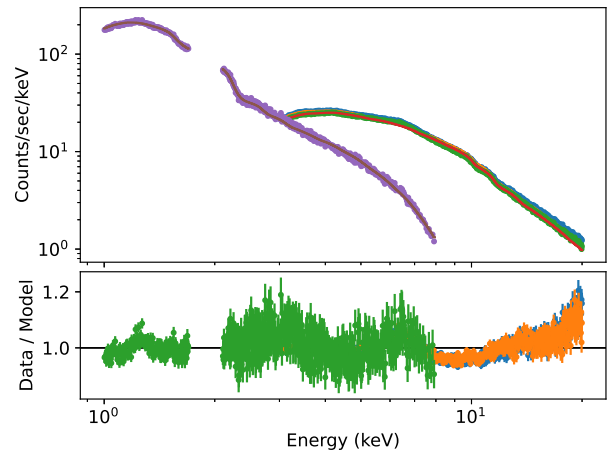}
\caption{Cyg\,X-1 {\it Suzaku} (blue data points),
{\it NuSTAR} (green and orange data points) data with the 
best-fit wedge-shaped corona model (lines).
The top panel shows the detection rate, and the lower
panel the ratio of the observed count rate divided by the
modeled count rate. 
\label{f:bf0}
}
\end{figure}

\begin{figure*}[t!]
\gridline{
\fig{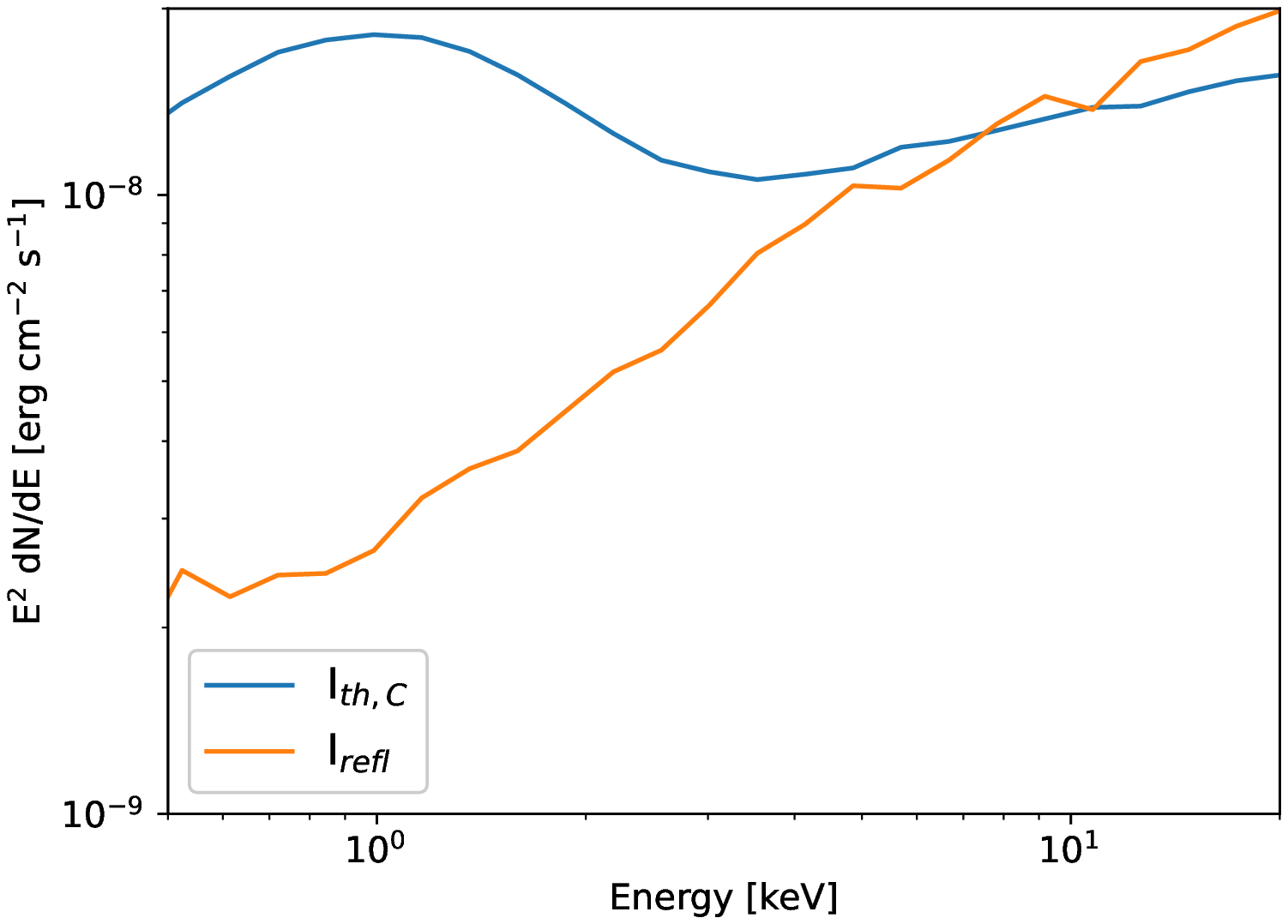}{0.45\textwidth}{(a)}
\fig{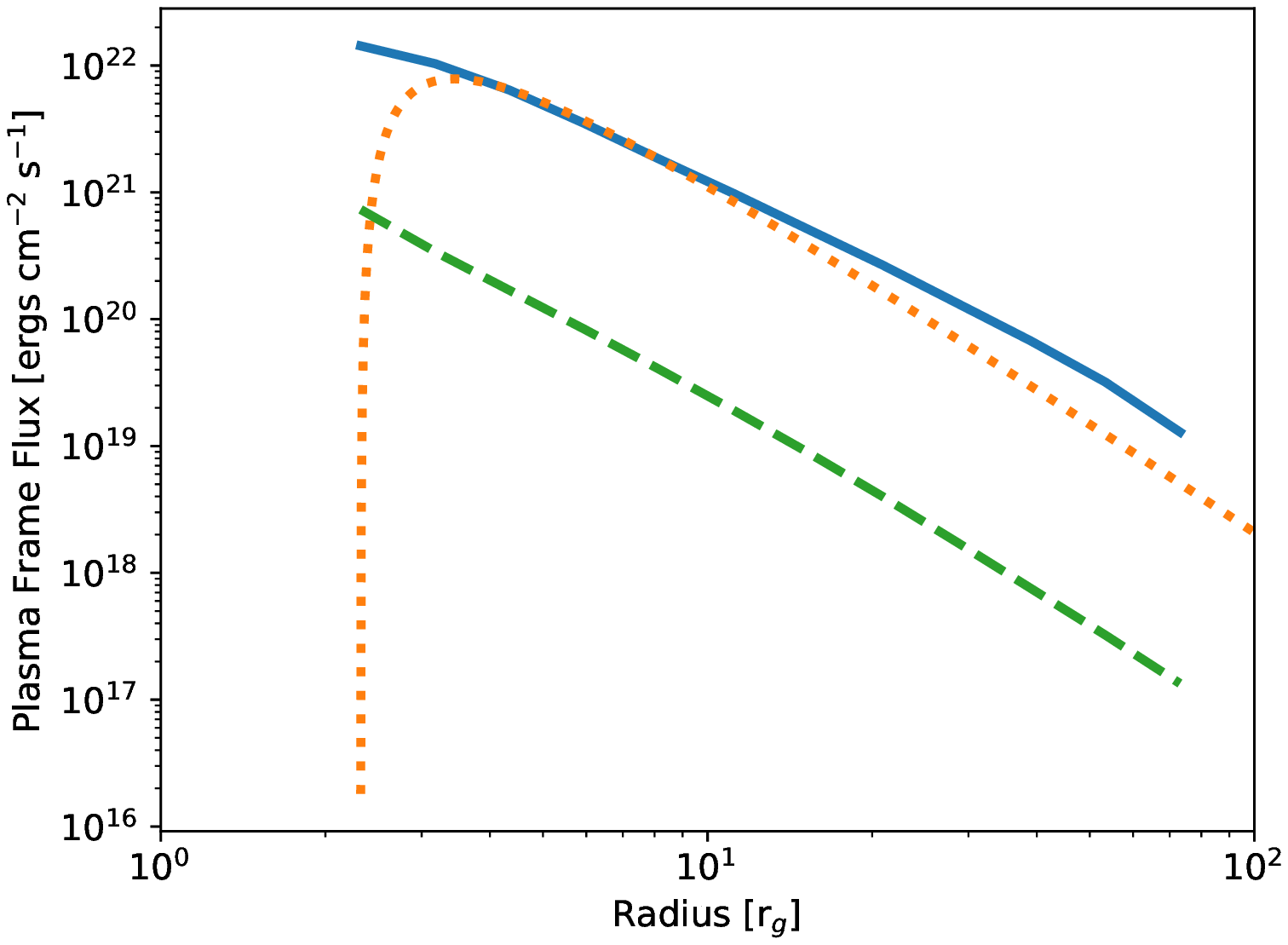}{0.45\textwidth}{(b)}}
\vspace*{-2ex}
\gridline{
\fig{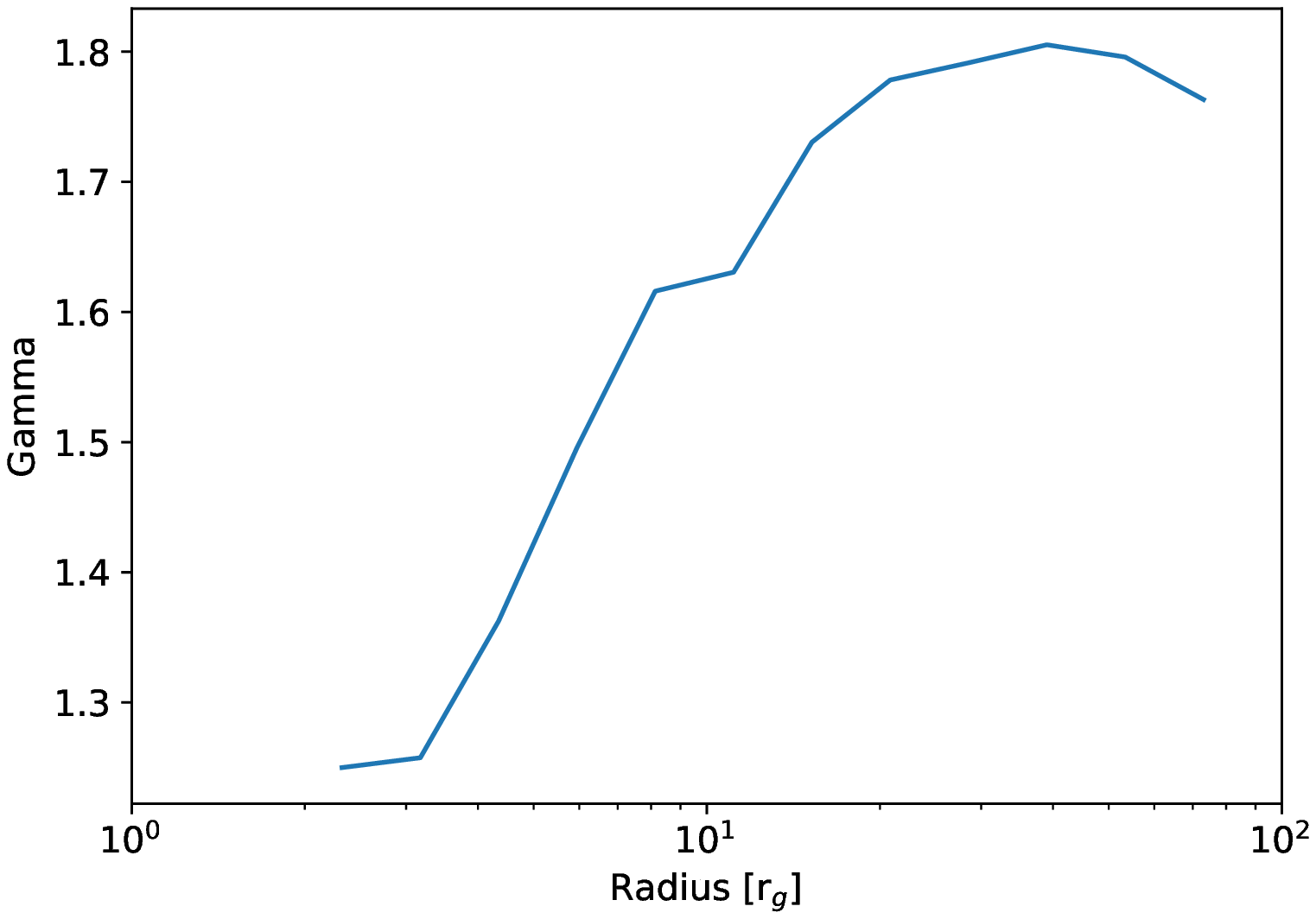}{0.45\textwidth}{(c)}
\fig{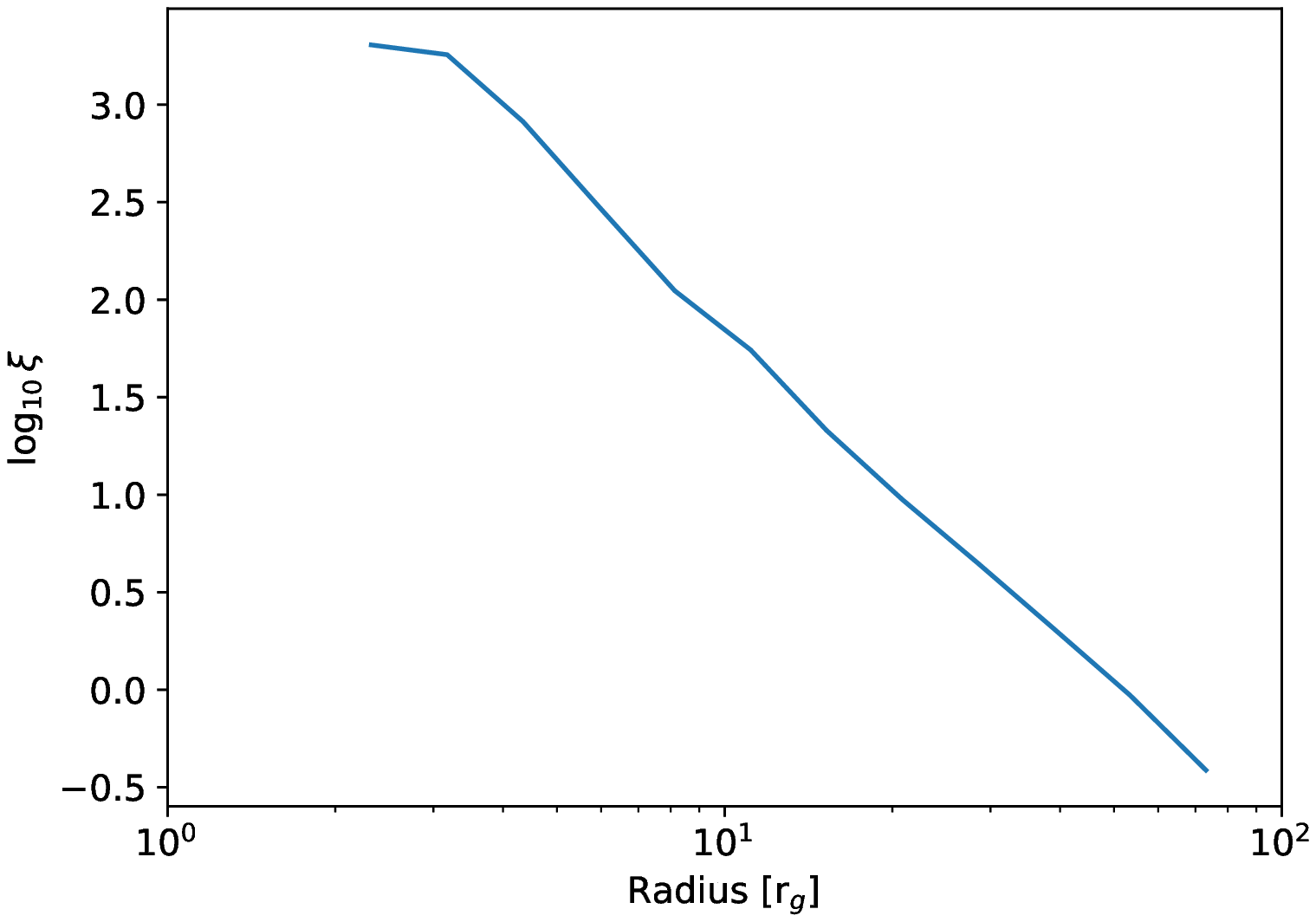}{0.45\textwidth}{(d)}}
\vspace*{-2ex}
\gridline{
\fig{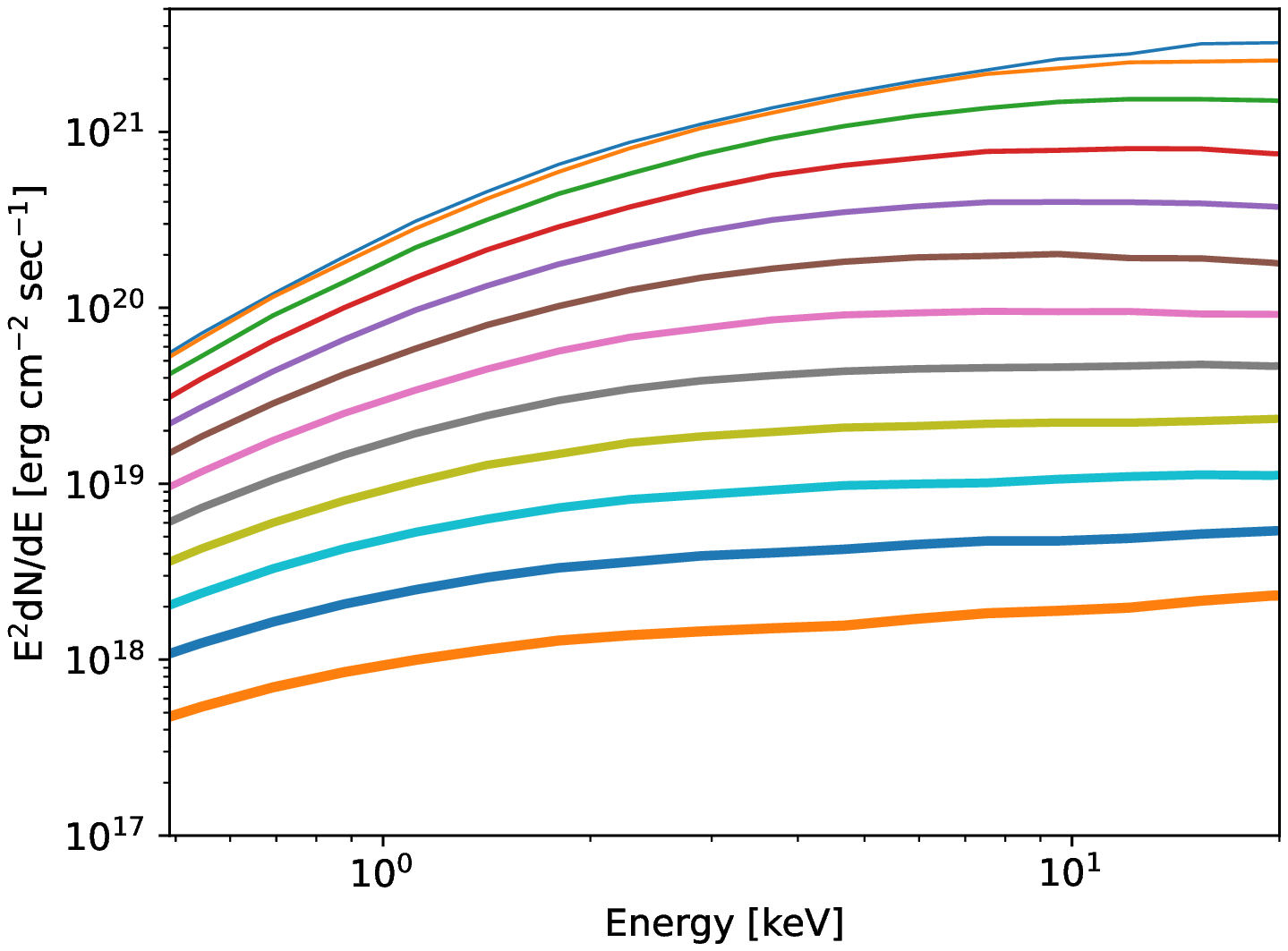}{0.45\textwidth}{(e)}
\fig{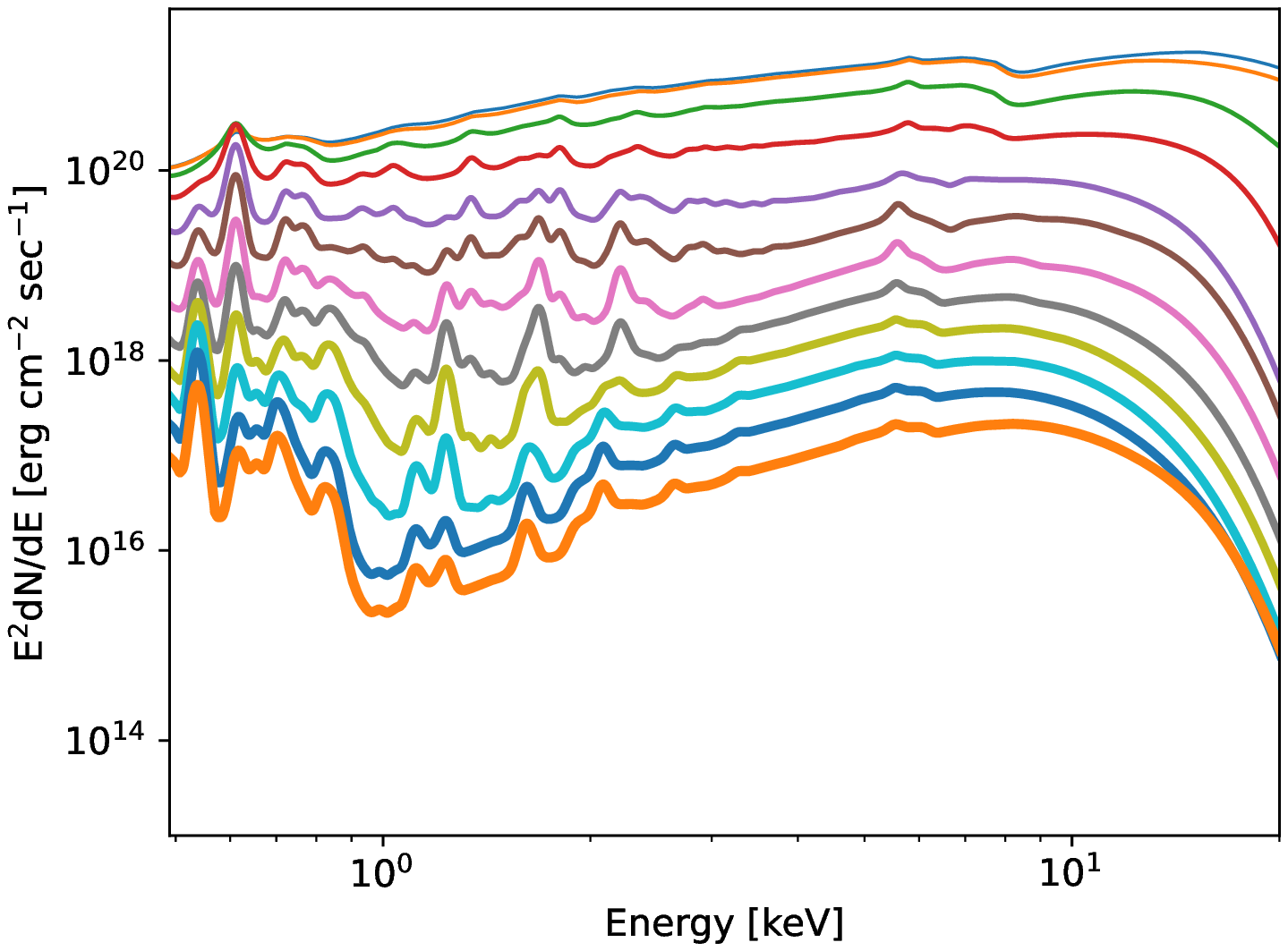}{0.45\textwidth}{(f)}}
\caption{{\tt kerrC} results for one of the configuration 
nodes of the best-fitting wedge-shaped corona model shown in Fig.\,\ref{f:bf0}
($a\,=\,0.9$,
$T_{\rm C}\,=\,100$\,keV, 
$\tau_{\rm C}\,=\,0.5$,
$\theta_{\rm C}\,=\,45^{\circ}$, and
$r_{\rm C}\,=\,100\,r_{\rm g}$).
(a) Thermal and coronal flux and reflected flux; 
(b) 0.1-1000\,keV plasma frame flux of the 
thermal disk emission (orange dotted line), 
the returning emission irradiating the disk (dashed green line),
and the returning and coronal emission (solid blue line);
(c) 1.5-15 keV photon indices of the emission
irradiating the accretion disk; 
(d) ionization parameter.
(e) plasma frame energy spectra of the returning and coronal 
emission irradiating the accretion disk for 12 radial bins with bin centers 
ranging from $r\,=\,2.3\,r_{\rm g}$ (thinnest line) 
to $73\,r_{\rm g}$ (thickest line).
(f): same as (e) but multiplied with the correction factor
$r$ from Equation (\ref{e:r}) for the inclination of
emission $\theta\,=\,41.4^{\circ}$. 
These energy spectra show the shape of the reflected
energy spectra.
{\tt kerrC} chooses for each individual photon the 
correction factor $r$ according to the photon's reflection angle.
\label{f:ir0}
}
\end{figure*}

The parameters of the best-fit wedge-shaped corona are listed in Table\,\ref{t:bf0}.
The model has a modest black hole spin of $a=\wa$, lower than the previously published values
from the fitting of the thermal state 
$a\,>\,0.9985$ \citep{2021Sci...371.1046M} and from similarly high values from 
fitting of the broadened Fe\,K-$\alpha$ line shapes \citep{2014ApJ...780...78T,2016A&A...589A..14D,2016ApJ...826...87W,2018ApJ...855....3T}. 

The wedge-shaped corona has a half opening angle of \wthetaC$^{\circ}$ and extends from  the ISCO at
$r_{\rm ISCO}\,=\,2.87$\,$r_{\rm g}$ to \wrCa\,$r_{\rm g}$. 
The corona temperature is \wtempC\,keV and the optical depth is \wtauC.
The photospheric electron density is $\log_{10}{n_{\rm e}}\,=\,$\wlogNe.
The {\tt XILLVER} electron temperature goes to the lowest possible value, indicating that the
fit prefers reflection energy spectra predicted for soft energy spectra irradiating the accretion disk.

Figure\,\ref{f:bf0} compares the best fit with the {\it Suzaku} and {\it NuSTAR} data.
Overall, the agreement is typically good to a few percent. 
The modeled 1-1.7 keV energy spectrum is missing a peak in the middle of this energy range.
Furthermore, the model does not fully reproduce the Fe\,K-$\alpha$ peak around 6.4\,keV, and is too soft at $>$15\,keV.

Figure \ref{f:ir0} gives some of the internal {\tt kerrC} results for the first configuration node
chosen by the interpolation engine ($a\,=\,0.9$,
$T_{\rm C}\,=\,100$\,keV, $\tau_{\rm C}\,=\,0.5$, $\theta_{\rm C}\,=\,45^{\circ}$, and
$r_{\rm C}\,=\,100\,r_{\rm g}$). This node usually impacts the results most strongly. 
Panel (a) shows the emission that did not experienced a reflection (blue) 
and the emission reflected at least once (orange). 
Both of these components may or may not have experience one or several Compton 
scatterings in the corona. The reflected emission starts to dominate at and above 8\,keV. 
Around $\sim$1\,keV, the flux without reflection is almost one order of magnitude stronger
than the reflected emission. Panel (b) shows the disk frame thermal emission from the disk
(dotted orange line), the returning emission (dashed green line), and the returning and coronal
emission (blue solid line). The dashed green line was calculated for the same accretion disk
without a corona. Two interesting conclusions: (i) the thermal flux from the disk approximately 
equals the coronal flux. Some of the accretion disk photons are energized (Comptonized) 
in the corona, and come back, giving rise to a coronal energy flux 
similar to the original accretion disk energy flux; (ii) the photons 
returning to the disk owing to spacetime curvature alone 
(without scattering in the corona), make up only a small fraction of the
energy flux irradiating the disk. Panel (c) presents the 1.5-15\,keV photon indices of the 
photons irradiating the accretion disk, showing a graduate steepening further away from
the black hole. The ionization drops from $\log_{10}{\xi}$\,$\sim$2 close to the
{\it ISCO} to $\log_{10}{\xi}$\,$\sim$0 at 100\,$r_{\rm g}$.
Finally, the last two panels show the comoving energy spectra of the photons irradiating the disk
(Panel (e)) and the emission leaving the disk (Panel (f)) for different radial bins (see the Figure caption for additional details). 
Interestingly, the fit does not choose a combination of 
higher $n_{\rm e}$ and lower $\xi$ (which would give more 
pronounced lines), as the overall shape of the energy spectrum 
does not fit well for lower $\xi$-values. 

\subsection{The Best-Fit Cone-Shaped Corona Model}
\begin{figure}
\includegraphics[width=0.5\textwidth]{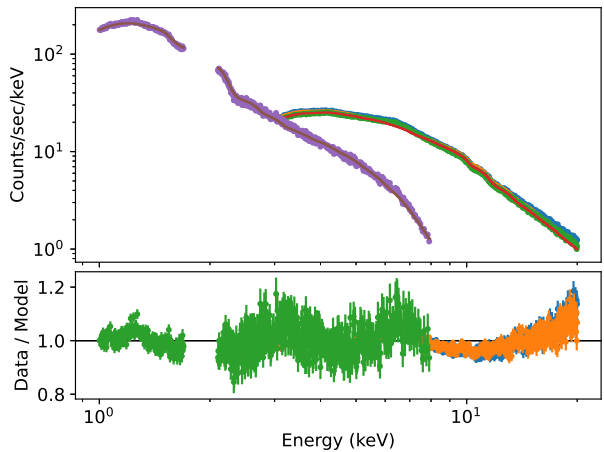}
\caption{Same as Fig.\,\ref{f:bf0} but for 
the best-fitting cone-shaped corona model.\label{f:bf1}
}
\end{figure}

The parameters of the best-fit cone-shaped corona are listed in Table\,\ref{t:bf1}.
This model has a slightly higher black hole spin of $a=\fa$ with
an innermost stable circular orbit at $r_{\rm ISCO}\,=\,2.17\,r_{\rm g}$. The corona is rather 
distant from the black hole and extends from \frCa\,$r_{\rm g}$ to  \frCb\,$r_{\rm g}$.
The opening angle is found at the largest simulated value of \fthetaC\,$^{\circ}$.
Overall, the corona is thus far away and rather extended.
The optical depth of the corona is with \ftauC\,larger than in the case of the wedge-shaped
corona (\wtauC), probably to provide the disk with a sufficiently 
large flux even though the corona is rather distant and lets
a good number of photons escape without scattering. 
The coronal temperature is with \ftempC\,keV slightly higher than 
in the case of the wedge-shaped corona (\wtempC\,keV). The photospheric electron 
density is $\log_{10}{n_{\rm e}}\,=\,16.3$, much lower than in the case 
of the wedge-shaped corona. The {\tt XILLVER} electron temperature is 50\,keV. 
Similar as for the wedge-shaped corona, the model fits the data within a few percent for most
of the range with larger deviations in the 1-1.7 keV energy range, around 6.4 keV, and at the highest energies (Fig.\,\ref{f:bf1}).

\begin{deluxetable}{p{3cm}p{2cm}}[t!]
\tablenum{4}
\tablecaption{Best-fit cone-shaped corona model. 
Only thawed parameters are listed; see Table \ref{t:bf0} 
for frozen parameters (including $n_{\rm H}$).
\label{t:bf1}}
\tablewidth{0pt}
\tablehead{
\colhead{Parameter} &  
\colhead{Result (90\% CL uncertainty)}}
\startdata
\multicolumn{2}{c}{Black Hole Parameters}\\ \hline
$a$  & \fa \\ \hline
\multicolumn{2}{c}{Accretion Flow Parameters}\\ \hline
$\dot{M}$  & \fMdot $\times 10^{18} \,{\rm g\, s}^{-1}$ \\
Corona radial extent
  & \frCa\,$r_{\rm g}$-\frCb\,$r_{\rm g}$ \\
$\theta_{\rm C}$  & \fthetaC$^{\circ}$\\
$kT_{\rm C}$  & \ftempC\,keV \\
$\tau_{\rm C}$  & \ftauC \\ \hline
\multicolumn{2}{c}{Reflection Parameters}\\ \hline
$kT_{\rm e}$  & \fkte\,keV \\
$\log_{10}n_{e,0}$   & \flogNe \\ \hline
\multicolumn{2}{c}{Fit Statistics}\\ \hline
$\chi^2/$DoF &  \fchi/\dof \\
\enddata
\end{deluxetable}

\begin{figure*}
\gridline{
\fig{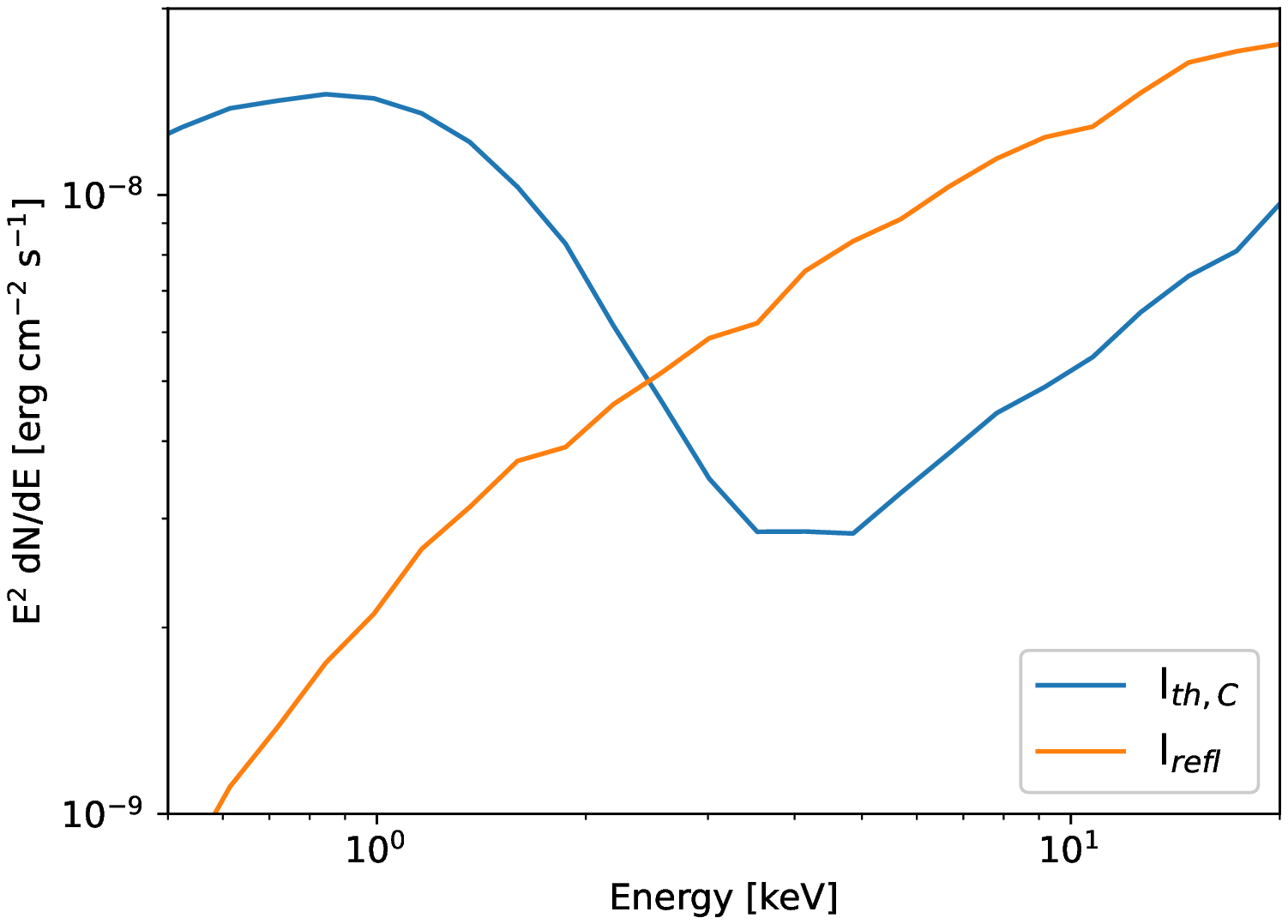}{0.45\textwidth}{(a)}
\fig{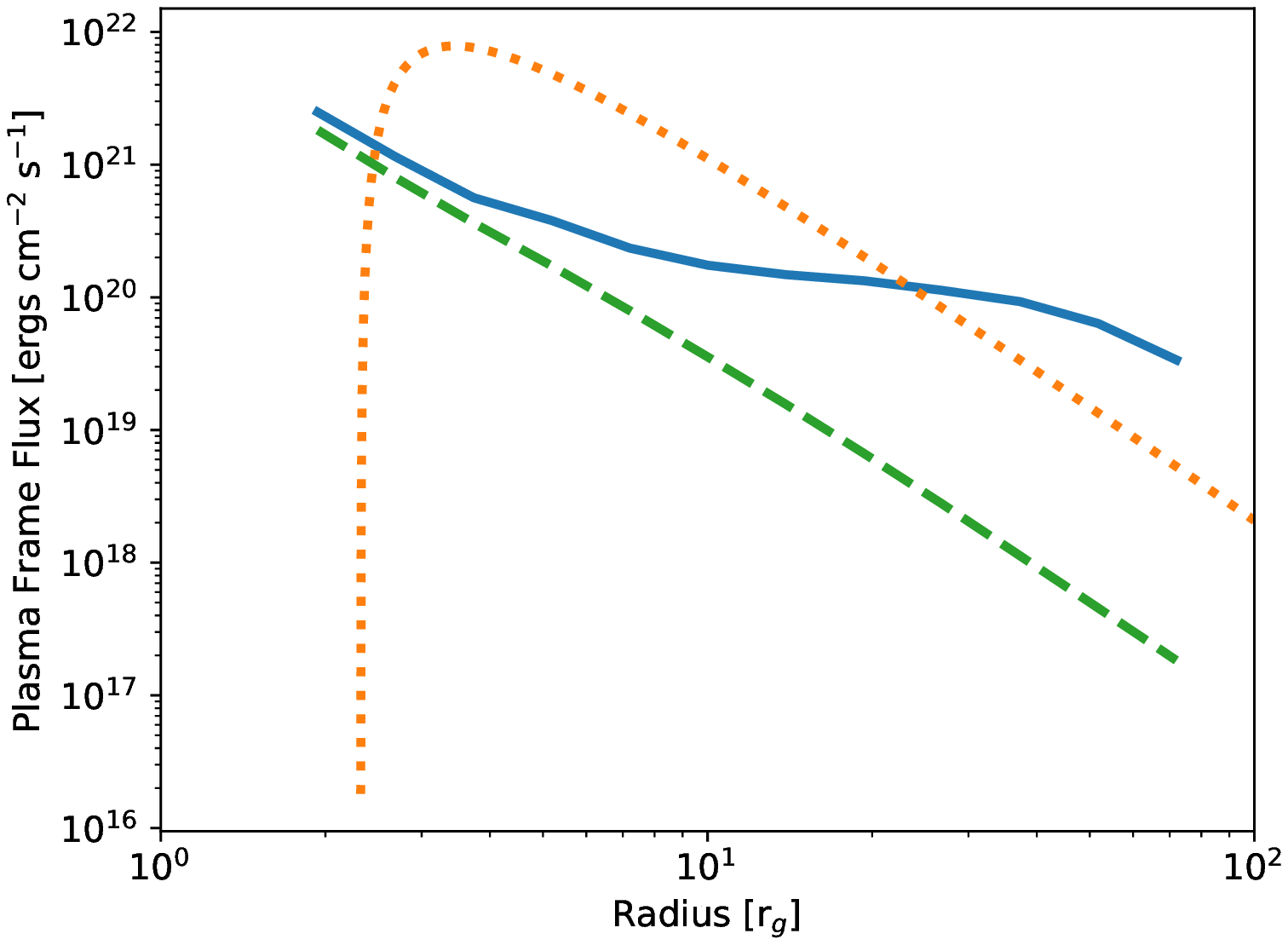}{0.45\textwidth}{(b)}}
\gridline{
\fig{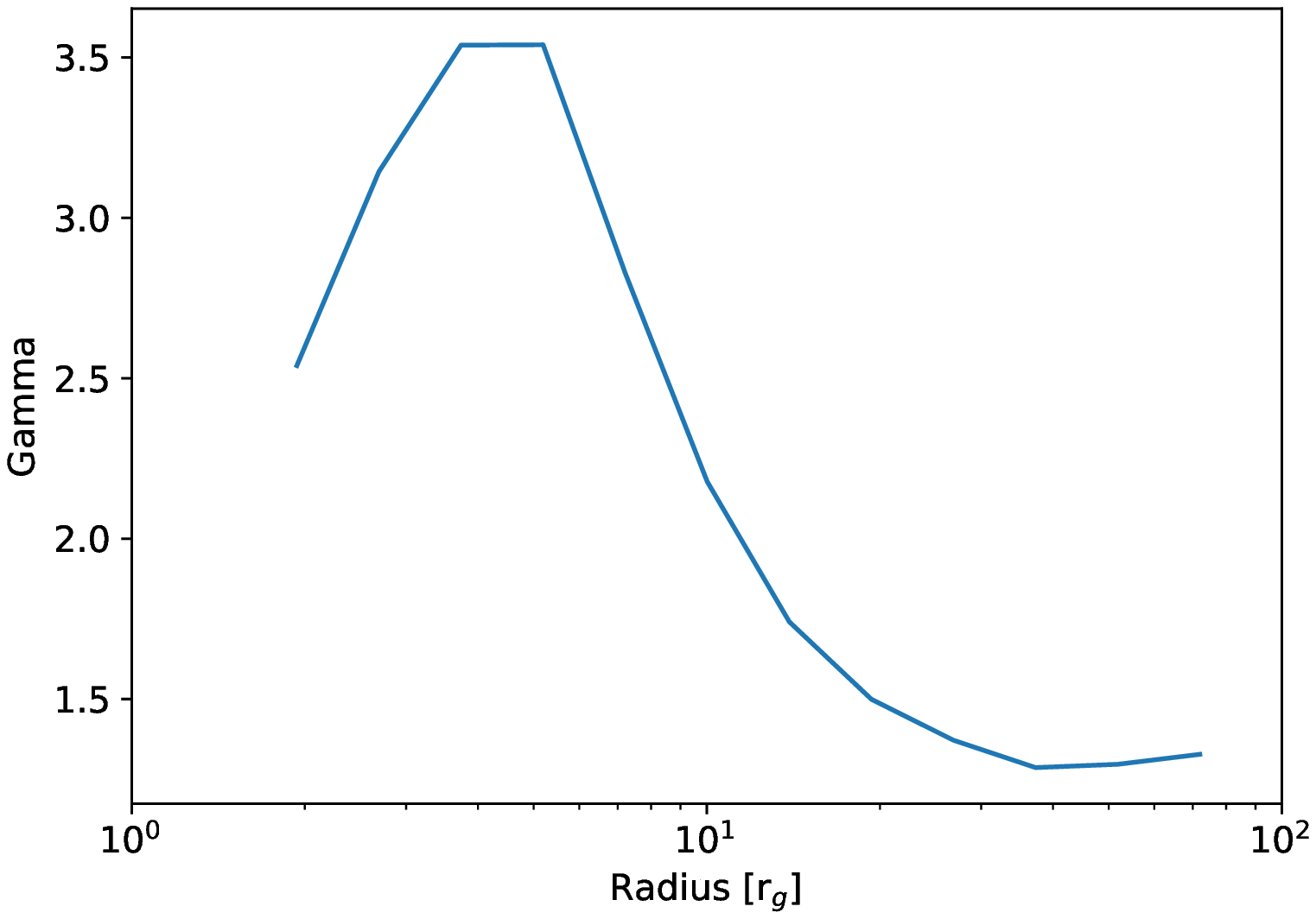}{0.45\textwidth}{(c)}
\fig{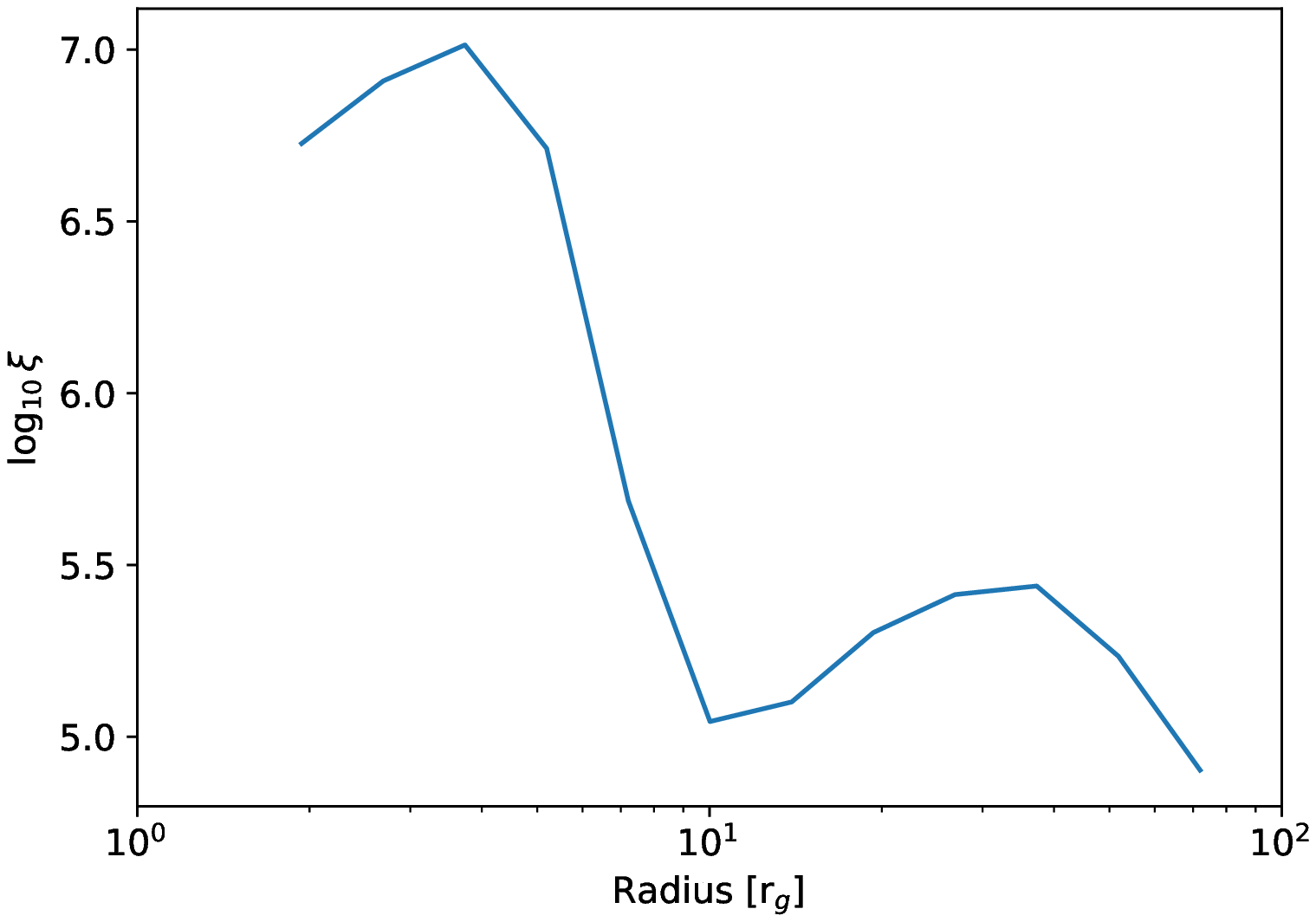}{0.45\textwidth}{(d)}}
\gridline{
\fig{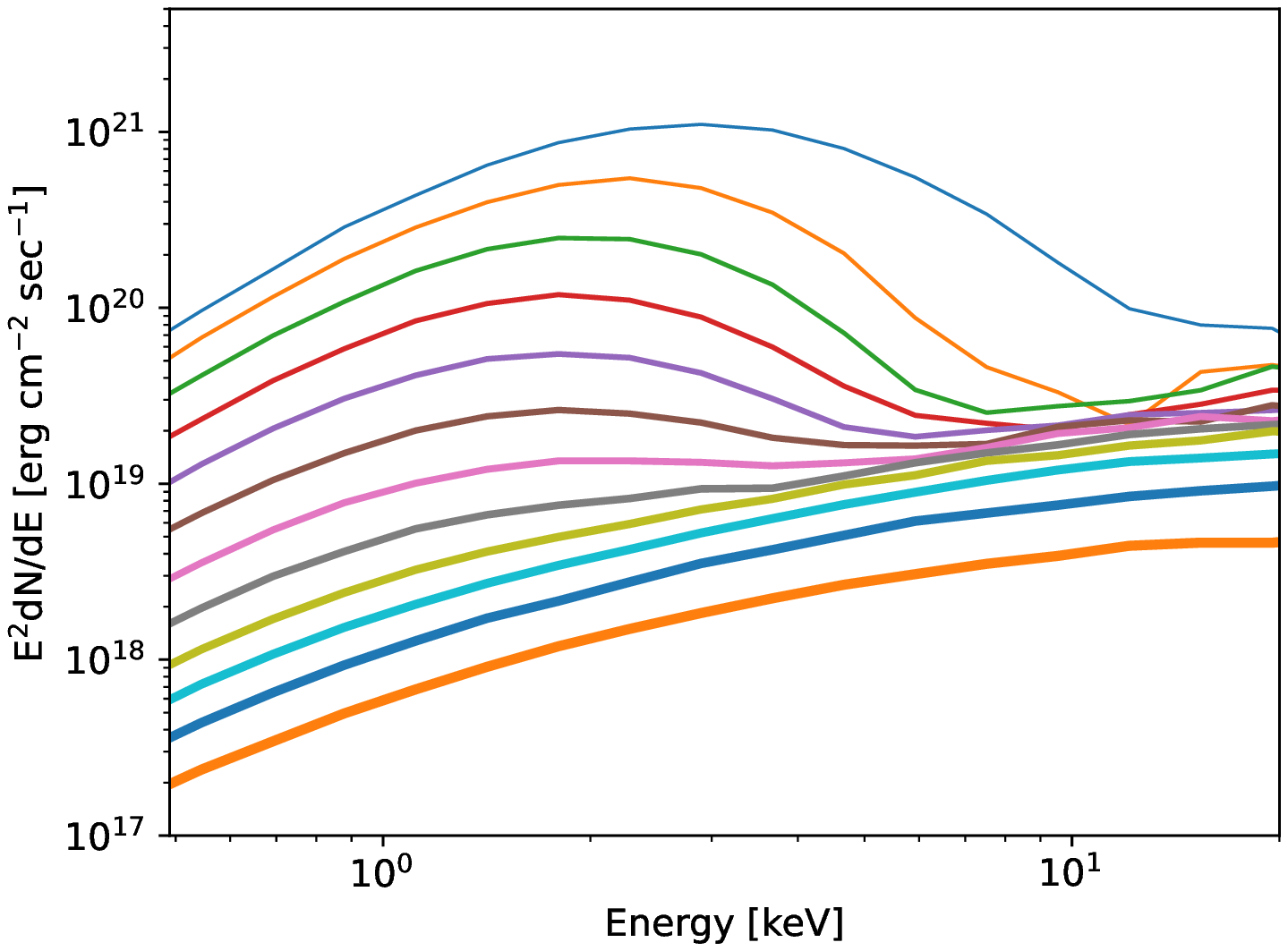}{0.45\textwidth}{(e)}
\fig{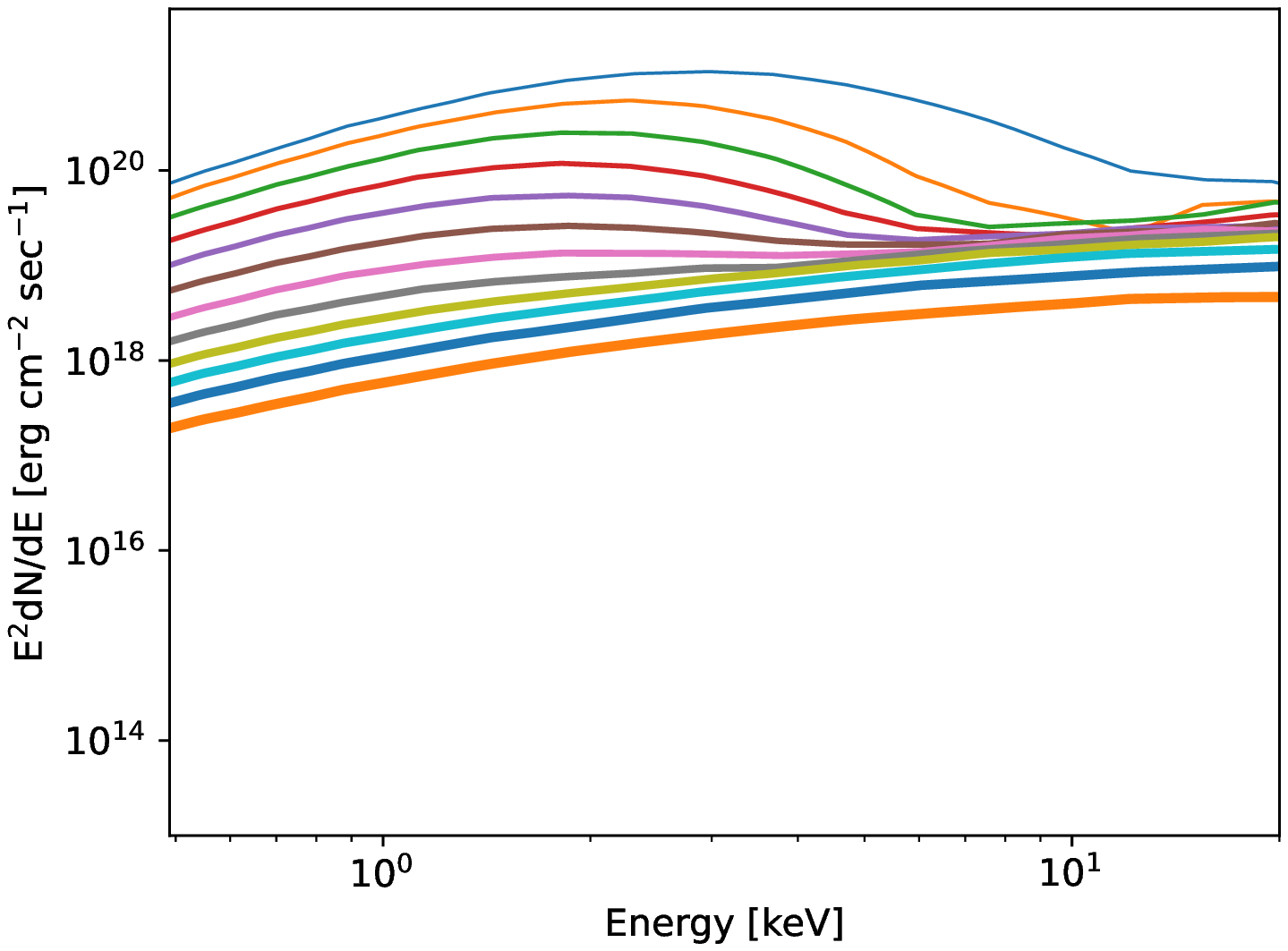}{0.45\textwidth}{(f)}}
\caption{Same as Fig.\,\ref{f:ir0}, but for 
one of the  configuration nodes of the best-fitting 
cone-shaped corona model 
($a\,=\,0.95$,
$T_{\rm C}\,=\,250$\,keV, 
$\tau_{\rm C}\,=\,0.75$,
$\theta_{\rm C}\,=\,45^{\circ}$, 
$r_1\,=\,50\,r_{\rm g}$, and
$r_2\,=\,100\,r_{\rm g}$).\label{f:ir1}
}
\end{figure*}

Figure \ref{f:ir1} shows detailed information for the first interpolation node 
($a\,=\,0.95$,
$T_{\rm C}\,=\,250$\,keV, 
$\tau_{\rm C}\,=\,0.75$,
$\theta_{\rm C}\,=\,45^{\circ}$, 
$r_1\,=\,50\,r_{\rm g}$, and
$r_2\,=\,100\,r_{\rm g}$). Panel (a) shows that the reflected energy spectrum 
starts to dominate over the non-scattered disk and coronal emission 
at much lower energies, i.e.\ around 2.5\,keV.
Panel (b) shows that the corona illuminates the accretion disk weakly at $r\sim r_{\rm ISCO}$,
where the returning thermal emission dominates the disk illumination.
The coronal Comptonized emission starts to dominate the disk illumination above $\sim$5\,keV.
Panel (c) presents the photon indices of the emission irradiating the disk. The energy spectra soften up to $\sim$4\,$r_{\rm g}$ where the returning emission dominates and harden at larger distances, 
where the coronal emission dominates. Panel (d) shows the resulting ionization parameter 
of the disk plasma. The pronounced minimum at $\sim10\,r_{\rm g}$ results from the combination of the
strong evolution of the energy spectrum as a function of the distance from the black hole
and the transition from the dominance of the returning radiation to the dominance of the
coronal emission.
Panel (e) shows how the energy spectra harden with distance from the black hole as
the coronal emission gains importance. Panel (f) shows that the reflected energy spectra 
are almost featureless close to the black hole and only show emission lines 
for the outermost radial disk bins.

The results indicate that cone-shaped coronae have difficulties Comptonizing a sufficiently large fraction of the accretion disk photons to account for the observed power law emission.
The best-fit corona thus resembles a large umbrella-shaped corona with a large opening angle,
enabling it to Comptonize a large fraction 
of the accretion disk photons. 
The fit chooses furthermore a low 
electron density and thus a 
very high disk ionization which leads to a 
high-yield of the reflected emission.
This result for the stellar mass black hole Cyg X-1 somewhat resembles similar difficulties
in some Active Galactic Nuclei (AGNs) where compact coronae close to their black holes
cannot explain the observed high reflected fluxes \citep{2016AN....337..441D}.   

\begin{figure*}
\gridline{
\fig{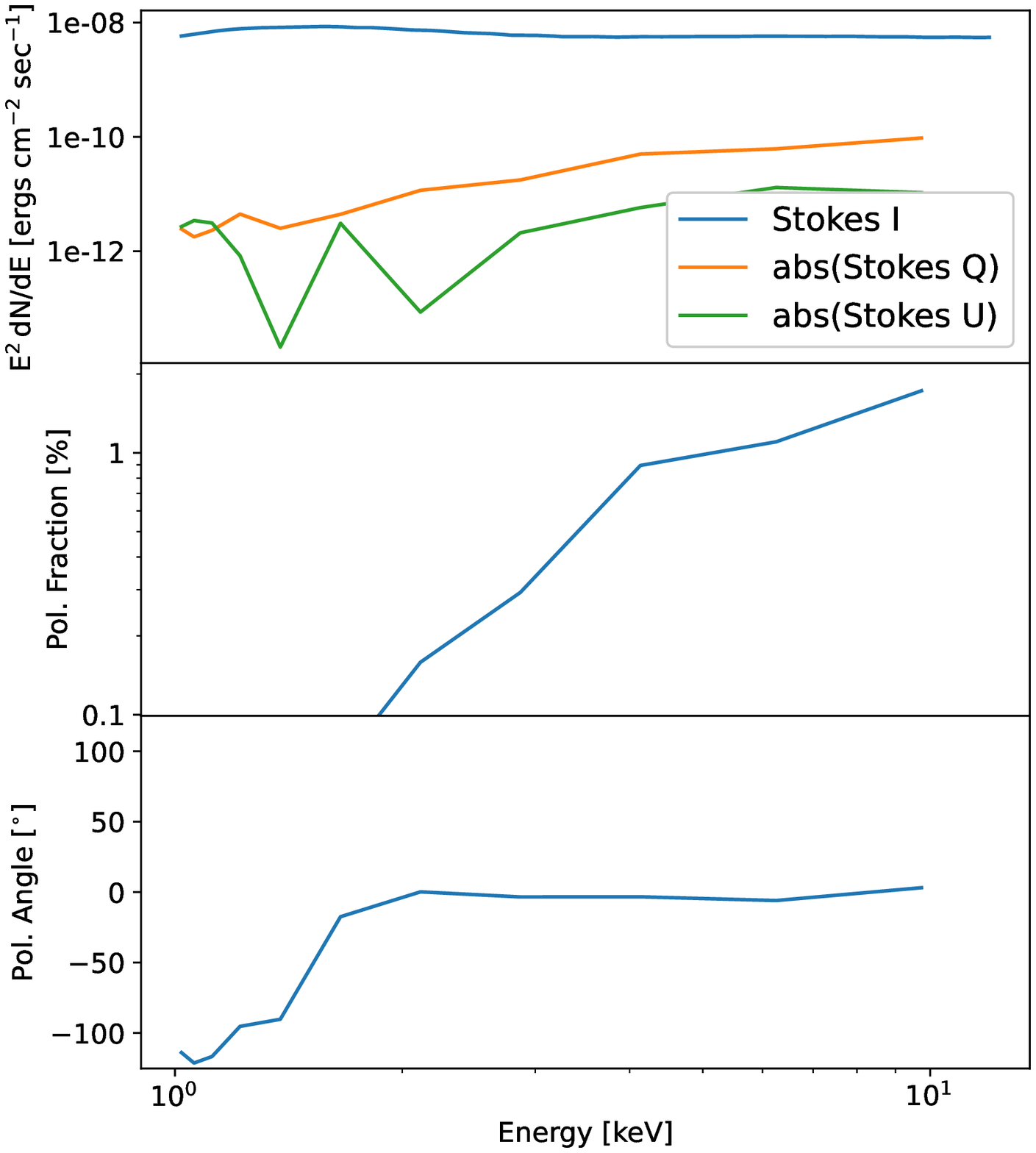}{0.45\textwidth}{(a)}
\fig{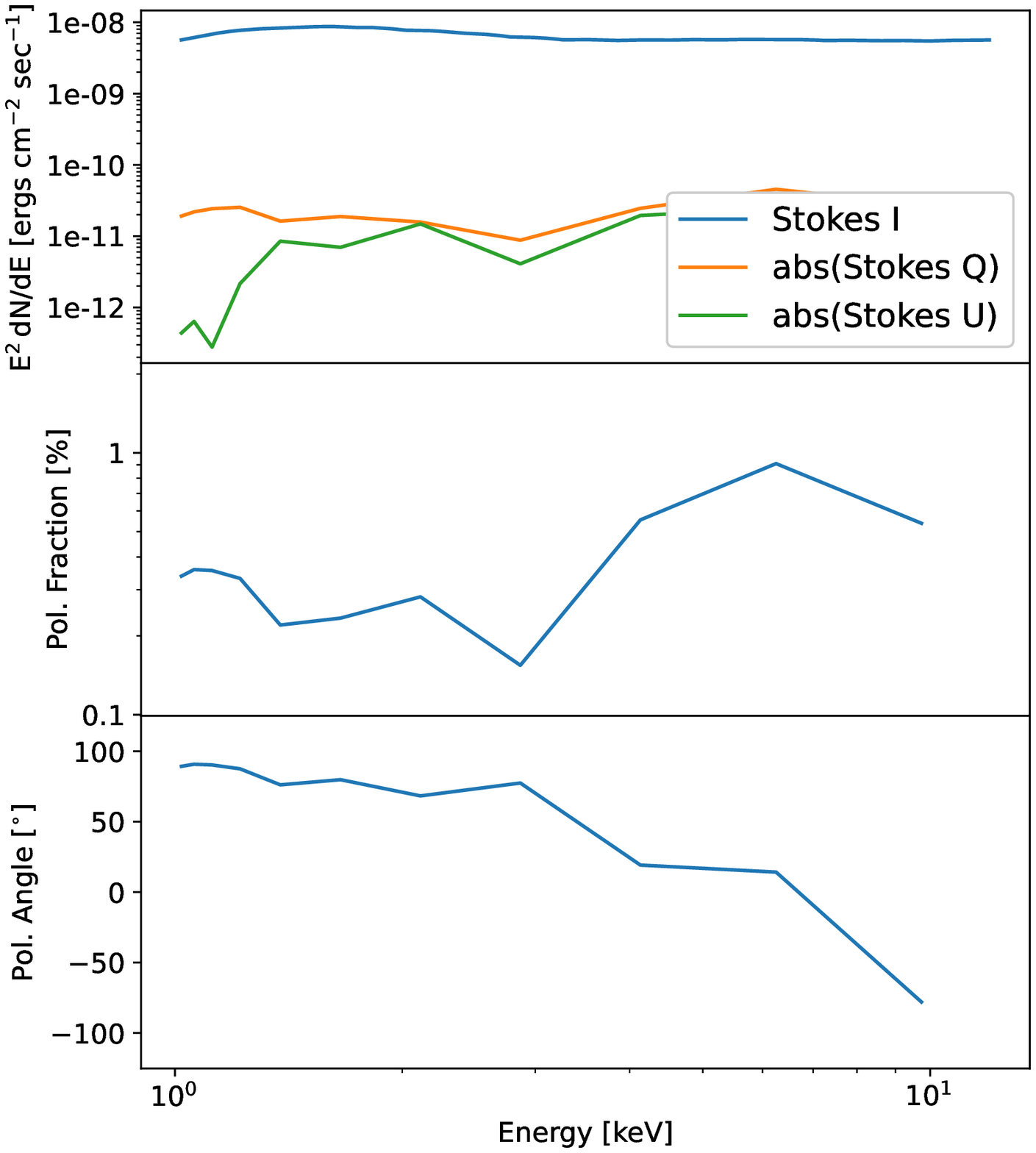}{0.45\textwidth}{(b)}}
\caption{Predicted $I$, $Q$, and $U$ energy spectra (top), polarization fractions (center), and polarization angles (bottom) for the best-fit wedge-shaped corona model (a) and the best-fit cone-shaped corona model (b).
Positive (electric field) polarization angles are measured counter-clockwise 
from the block hole spin axis. We are looking from above at 
an accretion disk rotating according to the right hand rule 
with the black hole spin pointing upwards. \label{f:p0}}   
\end{figure*}

\begin{figure*}
\gridline{
\fig{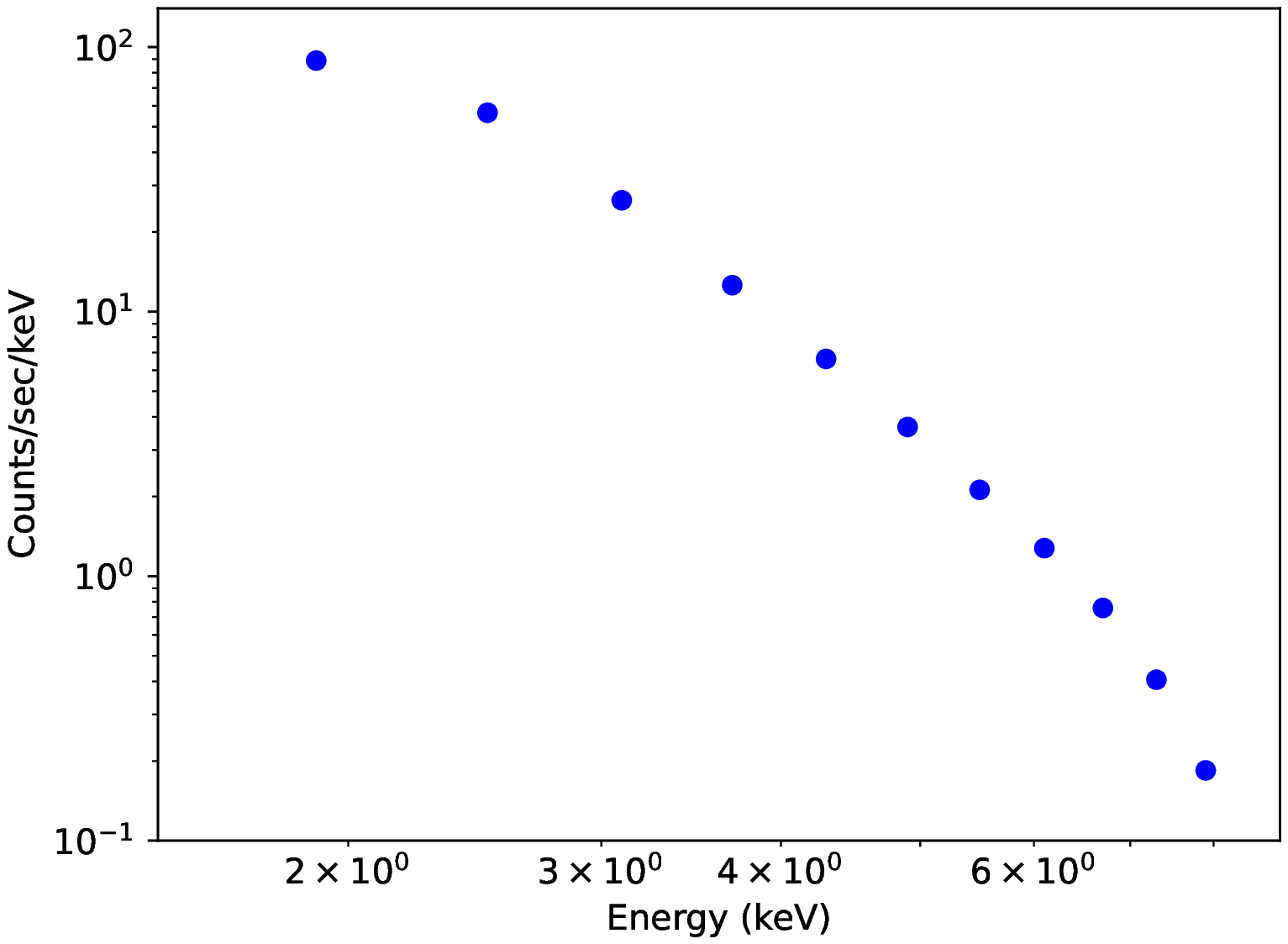}{0.45\textwidth}{(a)}
\fig{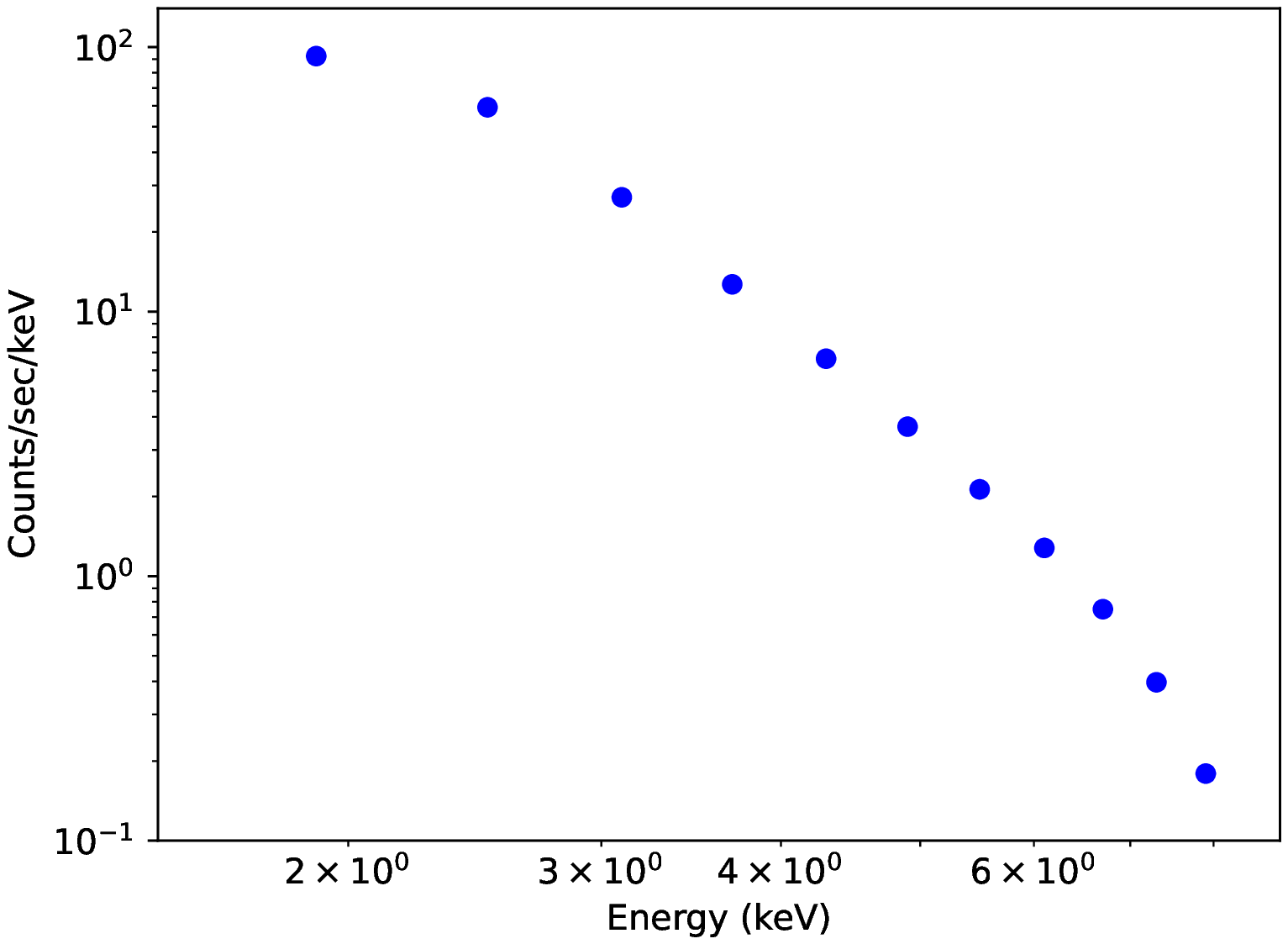}{0.45\textwidth}{(b)}}
\gridline{
\fig{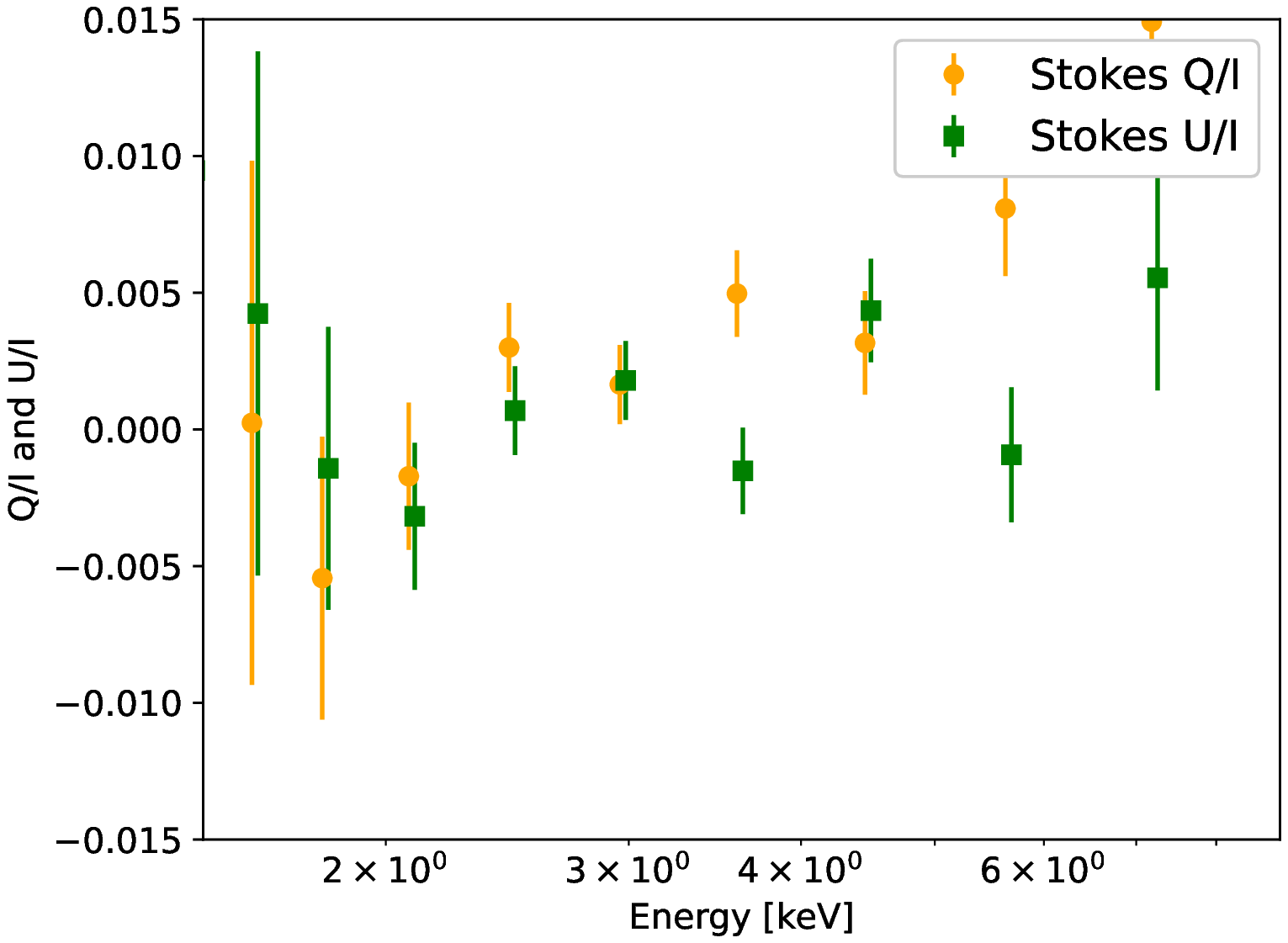}{0.45\textwidth}{(c)}
\fig{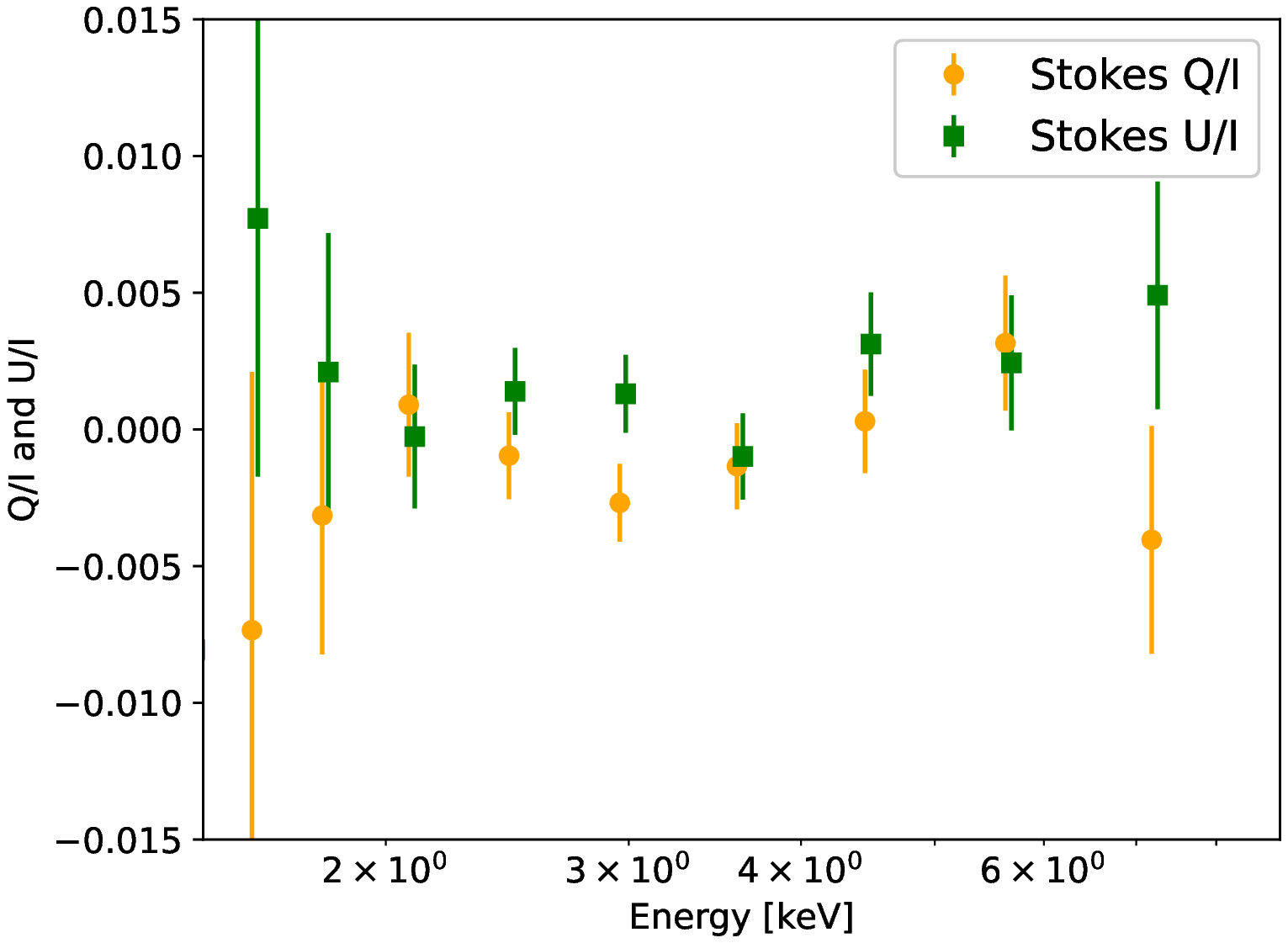}{0.45\textwidth}{(d)}}
\caption{Simulated outcomes of a 500\,ksec {\it IXPE} observation of Cyg\,X-1 for
the best-fit wedge model (a,c) and the best-fit cone model (b,d). The upper panels show Stokes $I$ and the lower panels show Stokes $Q/I$ and $U/I$. For Stokes $Q/I=1$ and $U/I=0$, the electric field of the X-ray emission 
is parallel to the black hole 
angular momentum vector;
for Stokes $Q/I=0$ and $U/I=1$, it is 45$^{\circ}$ rotated counter-clockwise. 
\label{f:p1}}
\end{figure*}

\section{Predicted {\it IXPE} Results}
\label{s:ixpe}
We developed a code to simulate and fit {\it IXPE} observations for
any model available in {\tt Sherpa}, including {\tt kerrC}. 
{\tt Sherpa} and {\tt X-Spec} user will know the {\it fake} command used to
combine a model with response matrices to generate simulated energy spectra.
Along similar lines, we added a {\tt Python} object {\tt simPol} that provides 
a {\it fake} method to generate Stokes $Q$ and Stokes $U$ energy spectra.

The code requires the model of the $I$, $Q$, and $U$ energy spectra, 
the Ancillary Response File (ARFs) (effective detection area as a function of energy), 
the Response Matrix Files (RMFs) (energy redistribution owing to detection principle 
and detector effects), and the Modulated Response File (MRF).
The MRF is the ARF times the energy-dependent modulation factor $\mu(E)$.
The modulation factor gives the fractional modulation of
the reconstructed polarization directions for a 100\% polarized signal 
and depends on the polarimeter performance,  and the event reconstruction methods.

The analysis of X-ray polarization data is based on assigning each detected X-ray photon a Stokes parameter $i=1$,$q\,=\frac{2}{\mu(E)}\cos{(2\psi)}$ and 
$u\,=\frac{2}{\mu(E)}\sin{(2\psi)}$ with $\psi$ 
being the reconstructed polarization direction 
and $E$ the reconstructed energy of the event 
\citep{2015APh....68...45K,2017ApJ...838...72S}. 
The sums $I$, $Q$, and $U$ of the $i$, $q$ and $u$ values 
of all events, respectively, 
are binned in energy and form the basis of the analysis.
The data are then analyzed by simultaneously fitting the detected and modeled 
$I$, $Q$ and $U$ energy spectra, using the uncertainties $\sqrt{I}$ on $I$,
and $\sqrt{2I}/\mu$ on $Q$ and $U$.
The {\tt simPol} {\it fake} method uses the $Q$ and $U$ standard deviations and 
$Q$-$U$ covariances to generate observed $Q$ and $U$ energy spectra  \cite[][]{2015APh....68...45K}.
Polarization fractions $p$ and directions $\chi$ can be calculated from 
$p\,=\,\sqrt{Q^2+U^2}/I$ and $\psi\,=\,\frac{1}{2}{\rm arctan2}(U,Q)$.
We use the preliminary {\it IXPE} ARFs, RMFs, and MRFs from \citet{Baldini2020}, and show
in the following only the results summed over all three {\it IXPE} detectors.

Figure \ref{f:p0} presents the polarization energy spectra 
for the two best-fit models for a simulated 500\,ksec {\it IXPE} observation
of Cyg\,X-1. The predicted polarization fractions turn out to be rather low for
the two models: lower than 1\% below 4\,keV and about 1\% at and above 4\,keV.
The $>$4\,keV polarization fractions is higher for the
wedge-shaped corona than for the cone-shaped corona.
In the 4-6\,keV energy range the expected polarization directions are 
parallel to the black hole spin axis.
For polarization fractions $<1\%$, the statistical errors on 
the predicted $Q$ and $U$ values are not entirely negligible.

The differences between $Q/I$ and $U/I$ energy spectra of the two models
come from the different black hole spins, the different polarizations 
that the photons acquire in the coronae of different shapes, and 
from the different accretion disk irradiation and reflection patterns.
Note that the reflected intensity depends among other factors
strongly on $\xi(r)$, which are very different for the two corona models. 

Figure \ref{f:p1} shows the simulated {\it IXPE} results for the two coronae.
For both corona models, the detection of a non-vanishing polarization
in the {\it IXPE} 2-8 keV energy range will be challenging. 
Note that the actual $Q$ and $U$ values depend on the orientation 
of the source in the sky. One could consider using a direction reference that aligns the
positive $Q$-axis with the axis of the VLBA jet \citep{2021Sci...371.1046M}. 
For the wedge-shaped corona, most of the signal will be expected in $Q$.
\section{Discussion}
\label{s:disc}
In this paper we describe a new X-ray fitting model, {\tt kerrC}, and its application to
intermediate state {\it Suzaku} and {\it NuStar} observations of the black hole Cyg\,X-1.
We chose an intermediate state observation as the thermal low-energy emission ($<3$\,keV) 
constraints the properties of the accretion disk. The power-law and line emission at higher 
 energies ($>$3\,keV) constrains the properties and location of the corona.  

Using {\tt kerrC} for the analysis of intermediate-state Cyg\,X-1 observations reveals the following findings:
\begin{enumerate}
\item A wedge-shaped corona above and below the accretion disk, and a cone-shaped corona in the funnel regions above and below the black hole describe the shape of the continuum energy spectra adequately, and can produce the observed relative intensities of the thermal and power law emission components. 
\item The wedge-shaped corona fits the data better than the cone-shaped corona. We included cone-shaped coronae in the funnel regions above and below the black hole in {\tt kerrC} as a possible 3-D approximation of the compact lamppost corona model. However, the fit chooses an extended cone-shaped corona with a large opening angle.  Even this large cone-shaped corona has difficulties to produce the observed hard emission, and the fit chooses a very high ionization fraction in order to maximize the yield of reflected photons. 
The high ionization fraction in turn gives a 
reflected energy spectrum almost entirely devoid of any emission lines.
\item Even the wedge-shaped corona underpredicts the relativistically broadened Fe K-$\alpha$ line.  The discrepancies may be reduced by adding model components, e.g., absorption or additional emission caused by material in the system \citep{2014ApJ...780...78T,2018ApJ...855....3T}. We did not explore these options in our analysis.
\item The thermal component constrains the black hole spin, and we obtain spin parameters between \wa and \fa, somewhat smaller than the results
of \citet{2016A&A...589A..14D,2016ApJ...826...87W,2018ApJ...855....3T,2021Sci...371.1046M}. 
It should be noted that the black hole spin and the location of the inner 
edge of the disk are somewhat degenerate model parameters. 
Neglecting the disk truncation can lead to underestimating the spin, as lower spins move
the radius of the innermost circular orbit outwards and can thus mimic disk 
truncations at $r>r_{\rm ISCO}$.
The code neglects the heating of the disk by the 
returning radiation and by the coronal emission.
Including this effect may lower the black hole spin estimate
as a hotter disk can mimic a disk extending to a smaller 
radius.
\item The energy spectra of the emission irradiating the accretion disk are not well described by simple Comptonized energy spectra but have rather complex shapes resulting from
the superpositon of returning emission, and  
partially  Comptonized coronal emission
(panels (e) in Figures \ref{f:ir0} and \ref{f:ir1}).
The result indicates that some of the assumptions
underlying inner disk line fitting are too simplistic.
The assumption of power law energy spectrum hitting the disk
will be a better approximation in the case of AGNs than in the case of stellar mass black holes, as in the former case 
lower energy accretion disk photons need a larger 
number of Compton scatterings before reaching X-ray energies.
The larger number of scatterings, and the larger ratio of the
energies of the coronal X-ray photons and the disk photons,
will both result in a cleaner power law energy spectrum in
the X-ray band.
\item For the small inclination of Cyg\,X-1, the expected polarization signal is small 
($\sim$1\% at 4-8 keV) but might be detectable with {\it IXPE}. 
These small polarization fractions are a bit smaller than the results
from the {\it OSO-8} observations of Cyg\,X-1 of 2.4\%$\pm$1.1\% at 2.6 keV 
and 5.3\%$\pm$2.5\% at 5.2 keV \citep{1980ApJ...238..710L}.
Note that the disk and corona geometries might evolve in time.
The state of Cyg\,X-1 during the {\it OSO-8} observations 
is poorly constrained.
\end{enumerate}

Possible improvements of {\tt kerrC} include the generation of a 
denser grid of simulated configuration nodes and larger numbers of events 
per node in the parameter regions inhabited by the actual astrophysical systems. 
It would be desirable to generate and use {\tt XILLVER} tables 
for input energy spectra that resemble the simulated energy spectra 
impinging on the disk more closely. 

It would be interesting to estimate the impact of the heating 
of the disk by the corona, and to improve the treatment of photons
returning many times to the disk. 

An interesting extension of {\tt kerrC} would include alternative geometries, 
including for example hot inner accretion flows in which the inner accretion 
disk disappears in favor of hot coronal gas, and seed photons enter 
the hot inner gas from a geometrically thin or thick accretion 
disk surrounding the hot central region.

{It would be desirable to implement the results of polarization dependent 
radiative transport calculations as in \citep{2020MNRAS.493.4960T,2021MNRAS.501.3393T,2022MNRAS.510.4723P}.
However, the Compton scattering cross sections 
and the scattering matrices for the reflection 
off the disk are polarization dependent. 
The approach adopted here of re-weighing 
pre-calculated geodesics would become much 
more complicated if the plasma properties 
(e.g. metallicity, ionization parameter) 
affect the polarization of the emitted and reprocessed photons.}

For the near future, our work will focus on using {\tt kerrC} to fit data from Cyg\,X-1 and other black holes for a range of
different emission states.
\section*{Acknowledgements}
The authors thank J. Tomsick for sharing the {\it Suzaku} data, and Luca Baldini for sharing the {\it IXPE} response matrices. 
HK thanks Javi\'{e}r Garc\'{i}a for the {\tt XILLVER} tables, for the collaborative work, and for detailed comments to the manuscript. HK thanks Michal Dov{\v{c}}iak, Wenda Zhang, 
Adam Ingram, and Giorgio Matt for the joint work in the {\it IXPE} stellar mass black hole working group.
We are grateful for the comments of an anonymous referee 
whose insights greatly benefited the paper.
The authors acknowledge the contributions of Quin Abarr (QA) and Fabian Kislat (FK) to the {\tt xTrack} raytracing code. 
QA developed and implemented the adaptive stepsize Cash-Karp integrator; FK implemented code changes that led to a major speed up of the code by avoiding the re-calculation of sine and cosine terms in the metric and in the Christoffel symbols.
HK thanks Andrew West, Arman Hossen,
Ekaterina Sokolova-Lapa, Mike Nowak,
Ephraim Gau, Nicole Rodriguez Cavero, 
Lindsey Lisalda, Sohee Chun,
and Deandria Harper for the joint work.
This research has made use of software provided by the Chandra X-ray Center (CXC) in the application packages CIAO and Sherpa.
HK acknowledges NASA support under grants 80NSSC18K0264 and NNX16AC42G.

\bibliography{main}{}
\bibliographystyle{aasjournal}

\end{document}